\documentclass[%
 reprint,
 amsmath,amssymb,
 aps,
pra,
superscriptaddress,
]{revtex4-1}

\usepackage{graphicx}
\usepackage{dcolumn}

\usepackage{amssymb}
\usepackage{amsmath}
\usepackage{bm}
\usepackage[utf8]{inputenc}
\usepackage{dsfont}
\usepackage[capitalize]{cleveref}
\usepackage{bbold}
\let\revappendix\appendix

\newcommand\dif{\mathop{}\!\mathrm{d}}
\newcommand{\pdagger}{{\phantom{\dagger}}}
\renewcommand{\vec}[1]{\bm{#1}}

\usepackage[dvipsnames]{xcolor}  
\definecolor{darkgreen}{rgb}{0.0, 0.5, 0.0}

\usepackage{url}

\usepackage{todonotes}

\def \Jeff {J_\mathrm{eff}}  
\def \Jsd  {J_{sd}}          
\def \Teff {{\bm{{\cal T}}_\mathrm{eff}}}  

\begin{document}

\preprint{}

\title{Nonequilibrium dynamics in a spin valve with noncollinear magnetization}

 \author{Rudolf Smorka}
 \affiliation{Institute of Physics, Albert-Ludwigs-Universität Freiburg, Hermann-Herder-Straße
3, 79104 Freiburg i. Br., Germany}

 \author{Pavel Bal\'a\v{z}}
 \affiliation{FZU -- Institute of Physics of the Czech Academy of Sciences, Na Slovance 1999/2, 182 21 Prague 8, Czech Republic}
 
 \author{Michael Thoss}
 \affiliation{Institute of Physics, Albert-Ludwigs-Universität Freiburg, Hermann-Herder-Straße
3, 79104 Freiburg i. Br., Germany}
\affiliation{EUCOR Centre for Quantum Science and Quantum Computing,
Albert-Ludwig University Freiburg, Hermann-Herder-Strasse 3, 79104 Freiburg, Germany}

 \author{Martin \v{Z}onda}
\affiliation{Department of Condensed Matter Physics, Faculty of Mathematics and Physics, Charles University, Ke Karlovu 5, Praha 2 CZ-121 16, Czech Republic}
\affiliation{Institute of Physics, Albert-Ludwigs-Universität Freiburg, Hermann-Herder-Straße
	3, 79104 Freiburg i. Br., Germany}

\date{\today}

\begin{abstract}

We utilize a hybrid quantum-classical equation of motion approach to investigate the spin dynamics and spin-transfer torque in a spin valve under bias voltage. We show that the interplay between localized classical magnetic moments and conduction electrons induces a complex effective exchange coupling between the magnetic layers. This leads to a declination of magnetizations from layers anisotropy axes even in equilibrium. Introducing a finite bias voltage triggers spin currents and related spin-transfer torques which further tilt the magnetizations and govern the relaxation processes of the spin dynamics. Analyzing different scenarios of the applied bias voltage, we show that symmetric and asymmetric voltage drops can lead to relaxation times of the spin dynamics that differ by several orders of magnitude at comparable charge currents. In both cases we observe resonant features, where the relaxation is boosted whenever the chemical potential of the leads matches the maxima in the density of the states of the spin-valve electrons.           
 
\end{abstract}

\maketitle

\section{Introduction}

Magnetic multilayer devices, where the exchange coupling between magnetic layers is suppressed by a nonmagnetic interlayer, e.g.,  spin valves or magnetic tunnel junctions~\cite{dieny1994giant,nanomagnetism2006}, have attracted a lot of attention from engineers and scientists working in different fields. Besides their direct applicability as, e.g., various types of sensors~\cite{nanomagnetism2017}, in magnetic recording systems~\cite{nanomagnetism2017,Barnas2015,bhatti2017spintronics} or in the broader context of spintronics~\cite{spintronics2021,Hirohata2020}, they also provide a rich and accessible theoretical as well as experimental platform for the investigation of important physical phenomena. 
For example, in recent years, multilayer devices played a crucial role in the study of spin-hall effect~\cite{Liu2011,Jungwirth2012,Zheng2017}, ultra-fast demagnetization~\cite{Battiato2010,Eschenlohr2013,Balaz2018,bergeard2020tailoring}, 
domain-wall dynamics~\cite{Bajpai2019,balaz2020domain}, various types of spin-transfer torques (STT)~\cite{Slonczewski1996,Slonczewski1999,Slonczewski2002,Ralph2008,nanomagnetism2017,balaz2013spin} 
and the interplay between electronic transport and dynamics of localized magnetic moments in general~\cite{Stiles_PRB:2002,Waintal_PRB:2000,Petitjean_2012:PRL}.
In addition, spin-torque oscillators based on magnetic vortices in spin valves or
tunnel junctions became a promising candidate for
neuromorphic computing systems~\cite{Torrejon_Nature:2017,Romera_Nature:2018,Kanao_PRApp:2019,grollier2020neuromorphic,song2020skyrmion}.

Because of the diversity of these devices, which spread from molecular valve systems up to bulk~\cite{dieny1994giant,Evers2020,Guo2019,devkota2016organic}, a multitude of different theoretical 
methods are used to rationalize their properties and predict new features. 
Arguably, the most popular ones are classical micromagnetic simulations~\cite{Leliaert2019} 
and atomistic spin dynamics~\cite{Skubic_JPCM:2008} based on the Landau–Lifshitz–Gilbert (LLG) equation. 
A clear advantage of these methods is the large number 
of highly optimized and versatile computer codes available~\cite{Vansteenkiste_AIPAdv:2014,Skubic_JPCM:2008},
which allow to address large systems and to incorporate experimentally measured parameters~\cite{Leliaert2019}. However, to model phenomena where transport plays a crucial role, the LLG equation must be extended by phenomenological or approximate torque and damping terms, which describe the effective influence of spin currents on the magnetization~\cite{Abert2019}. Because transport is inherently a quantum phenomenon and because these terms are in general influenced by the changing state of the spin valve, such a simplistic treatment can miss important physics, especially in the case of systems far away from equilibrium.

On the other hand, fully quantum-mechanical approaches that are able to capture the quantum nature of these devices are usually constrained to small systems, static magnetic configurations, short times, or rely on severe approximations~\cite{Evers2020,Guo2019,mondal2019quantum,mondal2021when}.

A natural compromise between completely classical and fully quantum-mechanical approaches present hybrid methods which combine both classical and quantum degrees of freedom~\cite{stock2005classical,ohe2006dynamics,salahuddin2006self,Onoda2006,Sayad_2015,Sayad_2016,Sayad_2016_epl,ellis2017,Xie2017,Smorka2020,Petrovic18,Nikolic2018handbook,Bajpai2019,elbracht2020topological,Suresh2020,Bajpai2020,swain2021skyrmions}. In the case of magnetic systems, these methods consider classical localized magnetic moments interacting with quantum conduction electrons. 
In their simplest form, a strict separation of time scales is assumed; that is, the dynamics of the classical spins is considered to be much slower than the one of electrons. Under this assumption, electrons respond instantaneously to the slow time-dependent potential of the classical degrees of freedom and, therefore, can be described by steady-state approaches, e.g., via nonequilibrium Green functions (NEGF)~\cite{ohe2006dynamics,salahuddin2006self,ellis2017,Xie2017,Zonda2019,Smorka2020}.
However, several recent studies have shown that this approach is often invalid~\cite{Sayad_2015,Sayad_2016,Sayad_2016_epl,Petrovic18,elbracht2020topological,Nikolic2018handbook,Bajpai2019,Suresh2020,Bajpai2020} because the two time-scales can not be strictly separated in general. 

To take account of this issue, one has to resort to non-Markovian approaches, in which electrons react in a finite time to the changes of the classical spins~\cite{Sayad_2015,Sayad_2016,Sayad_2016_epl,Petrovic18,elbracht2020topological,Nikolic2018handbook,Bajpai2019,Suresh2020,Bajpai2020,smorka2021singlespin}. These approaches reveal a time-dependent misalignment between the localized magnetic moments and the local nonequilibrium spin density of conduction electrons. Its most important consequences materialize in the form of additional torques and time-retarded damping effects~\cite{Sayad_2015,Sayad_2016,Sayad_2016_epl,Petrovic18,elbracht2020topological,Nikolic2018handbook,Bajpai2019,Suresh2020,Bajpai2020}. 

Nevertheless, there are quantum effects not fully captured even by these methods. 
For example, they do not account for so-called quantum spin-transfer torque 
resulting due to the quantum many-body states~\cite{Petrovic2021} and Kondo effect~\cite{Hewson1993,Sayad_2015}, as neither is included in the effective single-particle picture of the hybrid methods. It is also questionable if they can describe the relaxation of large spins into an excited state due to the coherent coupling to reservoirs observed in quantum  systems~\cite{hama2018negative,hama2018relaxation,stegmann2020relaxation}, although there are some examples of nonthermal steady states in quantum-classical systems~\cite{eckstein2008nonthermal}. In the context of our study, it is also important to note that hybrid methods tend to underestimate the damping of the magnetic nutations~\cite{Sayad_2016_epl,neeraj2020inertial}. 

Despite these differences, the methods that combine classical localized spins with quantum conduction electrons are in a rather good qualitative agreement with the full quantum mechanical treatments~\cite{Sayad_2016,Sayad_2016_epl,elbracht2020topological,Petrovic2021} and capture most of its details.  
As such, these hybrid techniques proved to be extremely useful in the investigation of various phenomena not described by classical or adiabatic LLG based approaches, e.g., geometrical torque~\cite{elbracht2020topological}, magnetic inertia~\cite{Bajpai2019}, chiral spin and charge pumping ~\cite{Petrovic18}, formation of some nontrivial magnetic textures~\cite{Bajpai2020,Bostrom2019} or resonant dependence of the spin damping on voltage~\cite{filipovic2013spin,smorka2021singlespin}. 
In addition, they are generalizable to realistic band structures~\cite{Nikolic2018handbook}.

In this paper, we use a quantum-classical equations of motion (QC-EOM) approach for open quantum systems~\cite{Bajpai2019,smorka2021singlespin}, to 
study the dynamics of a spin valve system sandwiched between two metallic leads with finite voltage difference. QC-EOM is an Ehrenfest-type method \cite{stock2005classical,elze2012linear,bellonzi2016assessment} used, e.g., to study nuclear dynamics in quantum transport~\cite{verdozzi2006classical,metelmann2011adiabaticity} or current-induced bond rupture in single-molecule junctions~\cite{erpenbeck2018current}.  Its advantage is that in the case of noninteracting conduction electrons
the hierarchy is terminated exactly at the second tier for a general metallic band of the leads or at the first tier for the wide band limit (WBL) approximation~\cite{zheng2007time,jin2008exact,croy2009propagation,zhang2013first,popescu2016efficient,leitherer2017simulation}. 
The method is therefore numerically exact even far away from equilibrium, allows to reach long simulation times, and avoids approximate or phenomenological terms not resulting directly from the Hamiltonian of the model.

Using the QC-EOM method, we argue that the quantum character of the conduction electrons is crucial in understanding of the nonequilibrium dynamics of a spin valve. We show that this is true even in a seemingly simple case where the spin dynamics of the entire magnetic layer 
can be truthfully represented by a single aggregated macrospin. Moreover, the character of the voltage drop crucially affects its magnetic dynamics. The two commonly used types, namely, a finite voltage introduced by shifting the chemical potential of one lead and by an equal opposite shift of chemical potentials in both leads, show spin relaxation times that differ by several orders of magnitude at comparable charge currents.  

The rest of the paper is organized as follows. In section~\ref{sec:MaM}, we introduce the model and the QC-EOM method. In the results section~\ref{sec:Res}, we first discuss the isolated spin valve (Sec.~\ref{sec:isosv}). Here we introduce the macrospin approximation and show that the effective exchange coupling between magnetic layers is a complicated function of model parameters and system geometry reflecting the density of states of the conduction electrons. We then move to the driven system in Sec.~\ref{sec:Driven} where we initially discuss the transient dynamics, showing the crucial difference between various types of driving. Next, the relaxation of the magnetizations is analyzed and we show a staggering difference between symmetric and asymmetric voltage drop cases. Finally, we address current-driven torques and their effect on the steady state magnetization of the spin valve. Some technical aspects and derivations are provided in the Appendices. 

\section{Model and methods \label{sec:MaM}}

\begin{figure}[!b]
	\centering
	\includegraphics[width=1.0\columnwidth]{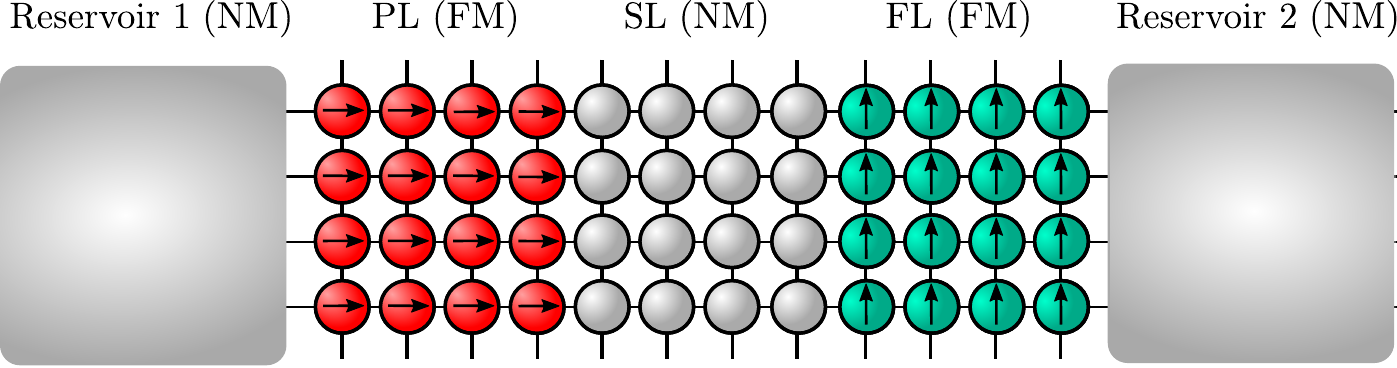}
	\caption{Schematic of a spin valve modeled on a two-dimensional square lattice which consists of ferromagnetic (FM) pinned 
	layer (PL or l, red spheres), nonmagnetic (NM) spacer layer (SL) and magnetic free layer (FL or r, wine spheres). Arrows depict localized classical spins $\{\vec{S}_j\}_{j\in\mathrm{PL,FL}}$. 
	The spin valve is coupled to two non-interacting metallic leads (gray area).
}\label{fig:SpinValveStructure}
\end{figure}

The spin valve heterostructure under consideration is illustrated in Fig.~\ref{fig:SpinValveStructure}. 
It consists of two ferromagnetic (FM) layers known as pinned (PL) and free layer (FL) separated by
a nonmagnetic (NM) spacer layer (SL)~\cite{dieny1994giant,camley2015magnetism}. 
We employ a hybrid quantum-classical description where localized spins 
are treated within a classical approximation but movable electrons are quantum particles.

The tight-binding Hamiltonian describing the interaction of quantum electrons with local time-dependent fields resulting from the interactions with localized spins on a lattice reads
\begin{align}
	 \bm{H}(t) =& -\gamma \sum_{\langle j,j'\rangle} \left(c_{j}^\dagger c_{j'}^\pdagger+\mathrm{h.c.}\right)+\overline{\mu}\sum_{j} c_{j}^\dagger c_{j}^\pdagger \nonumber \\
	 &+ \frac{\Jsd}{2}\sum_{j\in \textrm{PL},\textrm{FL}}c_{j}^\dagger \bm{\sigma}\cdot\vec{S}_j(t)^\pdagger c_{j}^\pdagger,
	 \label{eq:Hq}
\end{align}
where spinors $c_{j}=(c_{j\uparrow},c_{j\downarrow})^{T}$ and $c^\dagger_{j}=(c^\dagger_{j\uparrow},c^\dagger_{j\downarrow})$ consist of annihilation or creation operators of the conduction electrons with spins $\uparrow,\downarrow$ at site $j$. Their kinetic energy is described by the first term of the Hamiltonian. 
For simplicity, we set the electron nearest neighbor hopping parameter $\gamma$ constant in the whole spin valve. We use $\gamma$ as the energy scale, i.e., all energies presented in the text or figures are in the units of $\gamma$ and time is in units of $\gamma^{-1}$  (the typical range of $\gamma$ is $0.1-2$~eV \cite{Petrovic18,Bajpai2019,Suresh2020}).  
The second term of the Hamiltonian describes the influence of a constant electrochemical potential $\overline{\mu}$ which governs the electron occupation of the isolated spin valve. Here we assume that the system is small enough that its equilibrium electrochemical potential can be set and manipulated externally, e.g., by an auxiliary gate lead, which does not contribute to the charge and spin transport. 
If not stated otherwise, the electrochemical potential $\overline{\mu}$ is taken zero which in equilibrium or for isolated valve sets the half-filling condition.
The last term describes a local $sd$-like interaction between the electrons and classical spins with exchange coupling $\Jsd$. Here $\bm{\sigma}$ is the Pauli vector and $\vec{S}_j(t)$ is the magnetic moment vector localized at site $j$. 

The localized magnetic moments in the particular magnetic layer $\ell=\mathrm{PL,FL}\equiv l,r$ are described by the classical Hamiltonian
\begin{align}
H^\ell_\mathrm{C}(t) =& J_\mathrm{ex} \sum_{\langle j,j'\rangle} \vec{S}_j(t)\cdot\vec{S}_{j'}(t) 
- \sum_{j\in\ell} \vec{B}\cdot\vec{S}_j	\nonumber\\
&-K_\ell\sum_{j\in\ell} \left(\vec{S}_j(t) \cdot {\bm e}_\ell \right)^2 + 
\Jsd\sum_{j\in\ell} \vec{s}_j(t)\cdot\vec{S}_j(t),\label{eq:Hc}
\end{align}
where $J_\textrm{ex}$ is the intralayer exchange coupling between the neighboring localized magnetic moments, $\vec{B}$ is the vector of external magnetic field, $K_\ell$ is the layer-dependent magnetic anisotropy constant, while 
${\bm e}_\ell$ is a unit vector aligned with the local anisotropy easy axis. 
The last term couples the classical magnetic system to the quantum electrons. Here, vector $\vec{s}_j(t)=\frac{1}{2}\textrm{Tr}{\bm{\rho}_j(t)\bm{\sigma}}$ is the time-dependent electron spin-density, where $\bm{\rho}_j(t)$ is the reduced nonequilibrium single-particle density matrix of electrons on site $j$. All relevant physical constants have been absorbed into the parameters of the model and the magnitude of the localized magnetic moments (spins) is fixed to one. For simplicity, we address the structure described by the coupled Hamiltonians ~\eqref{eq:Hq} and~\eqref{eq:Hc} as the spin valve.

When considering an isolated system, that is, in the absence of fermionic reservoirs, the time evolution of the quantum part of the spin valve described by Hamiltonian~\eqref{eq:Hq} is governed by the Liouville-von Neumann equation for the single-particle electron density matrix $\bm{\rho}(t)$
\begin{align}
\frac{\partial \bm{\rho}(t)}{\partial t} &= -i\left[\bm{H}(t),\bm{\rho}(t)\right].\label{eq:EOM1}
\end{align}

In the presence of fermionic reservoirs, a system of equations of motion for the reduced density matrix $\vec{\rho}(t)$ is obtained by tracing out the reservoir degrees of freedom from the whole density matrix. In particular, to describe the dynamics of the magnetic junction, we use a hierarchical equations of motion approach \cite{tanimura2020numeric,batge2021nonequilibrium}. For the case of noninteracting fermions, studied in the present paper, the hierarchy of equations of motion for the auxiliary density matrices terminates at the second tier exactly \cite{zheng2007time,jin2008exact,zhang2013first,croy2009propagation,popescu2016efficient,leitherer2017simulation}. 

The equation of motion for the reduced single-particle density matrix reads
\begin{equation}
\frac{\partial}{\partial t}\bm{\rho}(t) = -i[\bm{H}(t),\bm{\rho}(t)] + \sum_\ell \left(\bm{\Pi}_\ell^\dagger(t) + \bm{\Pi}_\ell^{\phantom{\dagger}}(t)\right), \label{eq:EOM3}
\end{equation}
where the second term on the right hand side of \cref{eq:EOM3} generates dissipation, a nonunitary time evolution due to the coupling of the central system to the fermionic reservoirs. The current matrices $\bm{\Pi}_\ell(t)$ are expressed using the nonequilibrium single-particle greater/lesser Green functions
\begin{equation}
\bm{\Pi}_\ell(t)  = \int_{-\infty}^t\dif \tau \left[\bm{G}^> (t,\tau)\bm{\Sigma}^<_\ell(\tau,t)-\bm{G}^< (t,\tau)\bm{\Sigma}^>_\ell(\tau,t)\right].
\label{eq:AuxCur}
\end{equation}
Here, $\bm{\Sigma}_\ell^\lessgtr$ is the lesser/greater self-energy matrix due to the coupling between the reservoir $\ell$ and the spin valve. Assuming a constant density of states (WBL) in the reservoirs (leads) with constant broadening function $\Gamma_\ell$ (the matrix $\bm{\Gamma}_\ell$ has components $\Gamma_\ell$ at the interface $\ell$ and is zero otherwise), chemical potential $\mu_\ell$ and temperature $T=1/\beta$ we get   

\begin{align}
\bm{\Sigma}^<_\ell (t,\tau) &= i\int_{-\infty}^\infty \frac{\dif\varepsilon}{2\pi}\, f_\ell(\varepsilon,\mu_\ell,\beta)e^{-i\varepsilon(t-\tau)}\bm{\Gamma}_\ell,\label{eq:SE}\\
\bm{\Sigma}^>_\ell (t,\tau) &= -i\int_{-\infty}^\infty \frac{\dif\varepsilon}{2\pi}\, \left[1-f_\ell(\varepsilon,\mu_\ell,\beta)\right]e^{-i\varepsilon(t-\tau)}\bm{\Gamma}_\ell.\nonumber
\end{align} 
Here, $f_\ell(\varepsilon,\mu_\ell,\beta)$ is the Fermi function of reservoir $\ell$ which can be approximated by a sum over $N_p$ poles using the Padé representation~\cite{hu2010communication}
\begin{equation}
f(\varepsilon) \approx \frac{1}{2}-\frac{1}{\beta}\sum_{p=1}^{N_p}\eta_p \Bigg(\frac{1}{\varepsilon-\chi_{p\ell}^-} +\frac{1}{\varepsilon-\chi^+_{p\ell}}\Bigg),\label{eq:PadeDec}
\end{equation} 
where $\chi_{p\ell}^\pm = \mu_\ell\pm i\xi_p \beta^{-1}$ and $\eta_p$ are Padé coefficients. Employing the residue theorem, the above expansion allows to write the current matrices in an explicit form
\begin{equation}
\bm{\Pi}_\ell(t) = \frac{1}{4}(\mathds{1}-2\bm{\rho})\bm{\Gamma}_\ell+\sum_{p=1}^{N_p} \bm{\Pi}_{\ell,p}(t),\label{eq:AuxCur2}
\end{equation}
where the Padé-resolved auxiliary matrices $\bm{\Pi}_{\ell,p}$ follow the equations of motion
\begin{equation}
\frac{\partial}{\partial t} \bm{\Pi}_{\ell,p}(t) = -\frac{i\eta_p}{\beta_\ell}\bm{\Gamma}_\ell - i\left(\bm{H}-\frac{i}{2}\bm{\Gamma}-\chi_p^+\mathds{1}\right)\bm{\Pi}_{\ell,p}(t),\label{eq:EOM_AuxCur}
\end{equation}
with $\bm{\Gamma}=\sum_\ell\bm{\Gamma}_\ell$. Hence, within the wide band approximation used here, we get a closed (exact) system of EOM already at Eq.~(\ref{eq:EOM_AuxCur}). Note that the formalism is gauge invariant in the sense that the results for transport do not change if all three of the chemical potentials ($\mu_l$, $\mu_r$ and $\overline{\mu}$) are shifted by the same value.     

Finally, using the extension of classical Poisson-brackets to spin systems \cite{yang1980generalizations,lakshmanan2011fascinating}, the classical spin equation of motion for the magnetic moment at position $j$ reads 
\begin{align}
\frac{\partial \vec{S}_j(t)}{\partial t} &= \left\{\vec{S}_j(t),\bm{H}_\mathrm{C}(t)\right\}=
-\vec{S}_j(t)\times \nabla_{\vec{S}_j(t)}{H}_\mathrm{C}.
\label{eq:EOM2}
\end{align}

To obtain the overall time-dependence, we evolve the set of Eqs.~\eqref{eq:EOM2} together with Eq.~\eqref{eq:EOM1} for an isolated spin valve or, in the case of heterostructure, together with Eq.~\eqref{eq:EOM3} and Eq.~\eqref{eq:EOM_AuxCur} supplied by Eq.~\eqref{eq:AuxCur2}. In both cases, we evolve the system using the fourth-order (3/8-rule) Runge-Kutta  method with equal time steps for the quantum and classical subsystem. 

The current matrices $\bm{\Pi}_\ell(t)$ can be used to calculate the charge and spin currents between the spin valve and the leads $\ell={l,r}$
\begin{align}
{\cal I}_\ell(t) &= \pm\textrm{Re}\,\textrm{Tr}\left(\bm{\Pi}_\ell(t)\right),\label{eq:Cur}\\ {\cal J}^\alpha_\ell(t) &= \pm\textrm{Re}\,\textrm{Tr}\left(\left[\mathbb{1}_N\otimes\bm{\sigma}_\alpha\right]\bm{\Pi}_\ell(t)\right), \label{eq:SCur}
\end{align}
where the plus sign is for currents from the left ($l$) reservoir into the spin valve and minus for currents from the spin valve into the right ($r$) reservoir, $\bm{\sigma}_\alpha$ is the Pauli matrix and $N$ is the total system size (number of lattice points).  Similarly, the nonequilibrium single particle density matrix can be used to calculate the local charge and spin currents between particular monolayers~\cite{Wang2008,Nikolic2018handbook}. We pay special attention to the charge and spin resolved current at the SL-FL interface
\begin{align}
{\cal I}_\mathrm{F}(t) &=\frac{i}{2}\sum_{\left\langle j,j'\right\rangle \in \mathrm{IF}}\textrm{Tr}
\left[ \bm{H}_{j,j'}\bm{\rho}_{j',j} - \bm{\rho}_{j,j'}\bm{H}_{j',j}\right],\label{eq:Curl}\\
{\cal J}^\alpha_\mathrm{F}(t) &=\frac{i}{2}\sum_{\left\langle j,j'\right\rangle \in \mathrm{IF}}\textrm{Tr}
\left[\bm{\sigma}_\alpha\left(\bm{H}_{j,j'}\bm{\rho}_{j',j} - \bm{\rho}_{j,j'}\bm{H}_{j',j}\right)\right], \label{eq:SCurl}
\end{align}
where the sum runs over coupled pairs of nearest neighbors $\left\langle j,j'\right\rangle $ with $j$ taken from the last monolayer of SL and $j'$ from of the first monolayer of the FL. $\bm{H}_{j,j'}$ and $\bm{\rho}_{j,j'}$ are the respective $2\times 2$ submatrices of the quantum Hamiltonian and the nonequilibrium density matrix. The difference between the SL-FL interface spin currents ${\cal J}^\alpha_\mathrm{F}(t)$ and FL-lead interface spin current ${\cal J}^\alpha_\ell(t)$ can be used to enumerate the aggregated current-driven STT~\cite{Edwards2006}. However, because there can be a finite torque acting on the localized spins even in equilibrium, one has to subtract from the net torque equilibrium contributions to obtain the current-driven part of the STT
\begin{align}
\bm{{\cal T}}_\mathrm{cd}(t)=\bm{{\cal J}}_\mathrm{F}(t)-\bm{{\cal J}}_r(t)-\bm{{\cal J}}^\mathrm{eq}_\mathrm{F}.
\label{eq:torque}	
\end{align}  
	
To analyze nonequilibrium results, we also make use of the Landauer-B\"uttiker approach for the transmission function $\Theta$, its spin-resolved polarization $P$, and the density of states of the heterostructure $\mathrm{DOSh}$ ~\cite{haug2008quantum,Zonda2019} for a fixed configuration of classical spins $\bf{S}$ (typically the equilibrium one) 
\begin{align}
\Theta(\varepsilon,\bf{S}) &= \mathrm{Tr}\{ \bm{\Gamma}_l \bm{G}^R(\varepsilon) \bm{\Gamma}_r \bm{G}^A(\varepsilon)\},\label{eq:Tran}\\
P(\varepsilon,\bf{S}) &= \mathrm{Tr}\{ \bm{\Gamma}_l \bm{G}^R(\varepsilon) \left(\bm{\Gamma}^{\uparrow}_r-\bm{\Gamma}^{\downarrow}_r\right)  \bm{G}^A(\varepsilon)\}\label{eq:Pol},\\
\mathrm{DOSh}(\varepsilon,\bf{S}) & = \mathrm{Tr}\, i \{\bm{G}^R(\varepsilon) - \bm{G}^A(\varepsilon)\}/2\pi N\label{eq:Dosh},
\end{align}
where  $\bm{G}^{R(A)}$ is the retarded (advanced) Green function of the coupled system and $\bm{\Gamma}^{\sigma}_{l,r}$ are the spin-resolved coupling matrix between the system and the left ($l$) or right lead ($r$).

\section{Results \label{sec:Res}}


\subsection{Isolated spin valve}~\label{sec:isosv}
Before investigating the spin dynamics in an externally driven spin valve, it is instructive to first discuss the dynamics of an isolated spin valve. We use a one-dimensional case to discuss the effect of the electronic spectrum on the spin dynamics and the role of the nonmagnetic layer in the relaxation process, which both play an important role in the driven system.

\subsubsection{Electronic spectrum}
We first show that the details of the electronic spectrum significantly influence the magnetization dynamics of the spin valve. In general, the spectrum is sensitive to the orientation of the classical spins and acquires time dependence through their dynamics~\cite{smorka2021singlespin,filipovic2013spin}.  Already a simple system of just two localized spins coupled through spin-dependent currents can have very complicated dynamics, including some chaotic regimes~\cite{Onoda2006,lakshmanan2011fascinating}.  Therefore, to make our argument more comprehensible, we address here a case in which the electronic spectrum can be assumed to be mostly static and the dynamics of a particular spin in a spin valve is not too complicated with respect to its neighbors. 
\begin{figure}[!h]
	\centering
	\includegraphics[width=0.99\columnwidth]{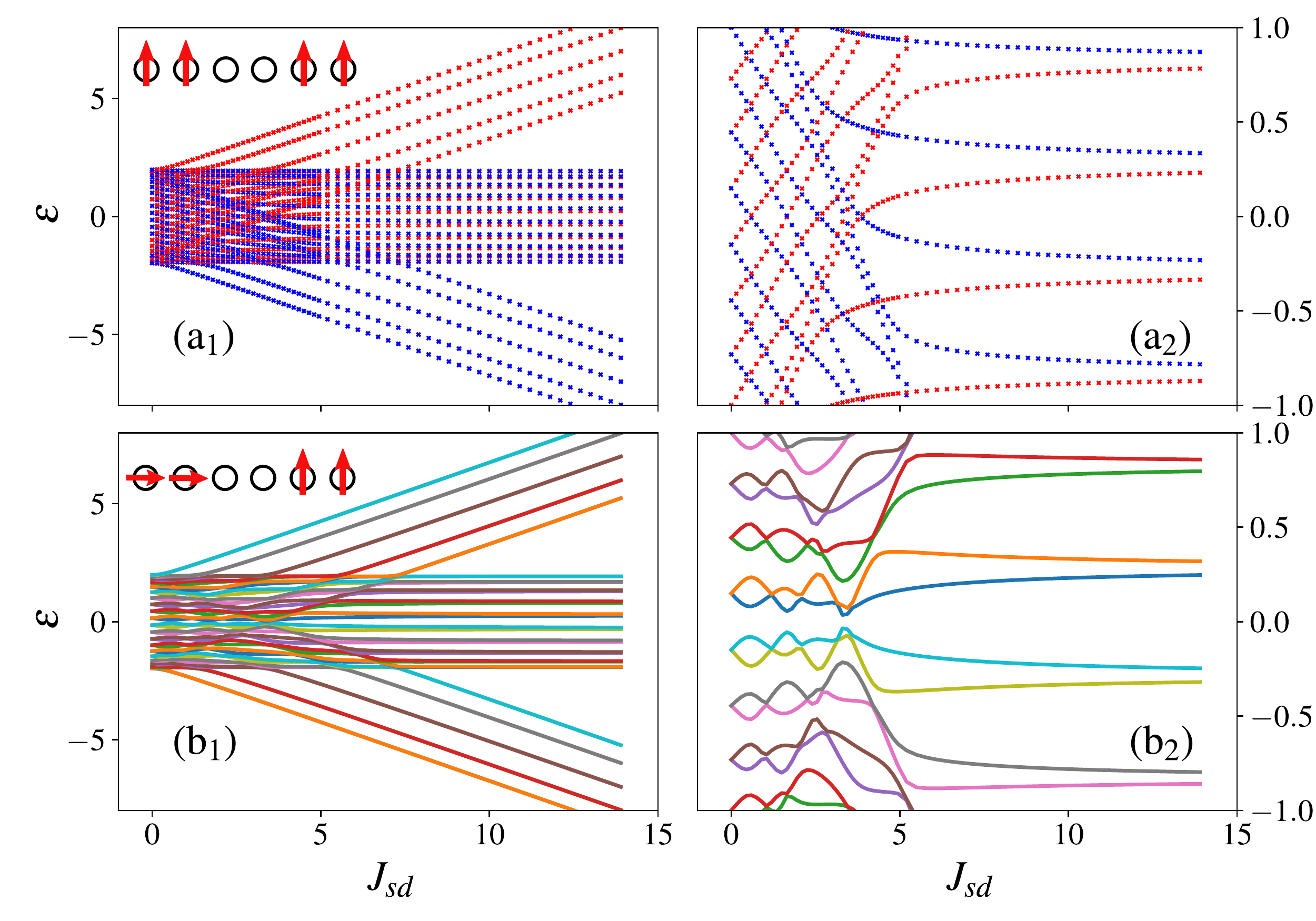}
	\caption{Electronic spectrum for a static one-dimensional layer with $N_{l(\mathrm{PL})}=5$, $N_\mathrm{NM}=10$, $N_{r(\mathrm{FL})}=5$ as a function of electron-spin coupling $\Jsd$. The two cases represent settings with ferromagnetic configuration within the layer and total normalized magnetization components within the layers being 
	(a) parallel: $M^z_l=1$ and $M^z_r=1$ where the two colors represent states with up (red) and down (blue) magnetic polarization,  
	(b) perpendicular $M^x_l=1$ and $M^z_r=1$. Panels on right show the details of the spectrum in the vicinity of the Fermi level.
	\label{fig:SpecTriLayer}}  
\end{figure}

We investigate a one-dimensional chain where the magnetic layers consist of one to ten sites each. We set the ferromagnetic Heisenberg exchange coupling $J_\textrm{ex}=-1$  and switch off the anisotropies $K_l=K_r=0$. This stabilizes a nearly ideal ferromagnetic ordering in both magnetic layers and significantly simplifies the dynamics. 

Before addressing the time evolution, it is useful to briefly discuss the electronic spectrum for some relevant static configurations of localized spins. Fig.~\ref{fig:SpecTriLayer} shows the dependence of the spectrum on $\Jsd$ for a linear system of total length $N=20$ ($N_\mathrm{NM}=10$) and two static configurations of the localized spins ($N_{l\mathrm{(PL)}}=N_{r\mathrm{(FL)}}=5$), a parallel one (a) and perpendicular one (b). 
Both spectra display a similar band splitting from one mixed band at small coupling ($\Jsd<3$) to three distinct bands in the strong coupling ($\Jsd > 7$) regime. Here, the top and bottom bands reflect the states predominately localized in the ferromagnetic layers (note their spin polarizations in Fig.~\ref{fig:SpecTriLayer}(a1) and the discussion in  \cref{App:loc}). 
Therefore, even when the presence of the central band can lead to seemingly finite DOS (under suitable broadening) at the Fermi-level for arbitrary $\Jsd$, the transport characteristics in the strong coupling regime can be still insulating-like. The reason is that the local DOS calculated for magnetic layers typically shows a large gap around the Fermi level. 
However, even in that case, the central band has an important influence on the spin dynamics, because the conduction electrons mediate an effective exchange interaction ($\Jeff$) between the magnetic layers. As we discussed below, $\Jeff$ is governed by the states in the vicinity of the Fermi-level. The real challenge is that even such a simple case as presented here shows a complicated $\Jsd$ dependence, including a rather complex avoided level crossing for weak coupling [Fig.~\ref{fig:SpecTriLayer}(b$_2$)].   

\subsubsection{Macrospin approximation}

The magnetic layers are coupled by an effective exchange interaction $\Jeff$ due to the presence of spin-polarized conduction electrons in the valve. Because of the strong exchange coupling $J_\mathrm{ex}$ which stabilizes the relative dynamics of spins within one layer, we can extract $\Jeff$ from the spin evolution by analyzing the dynamics of the net layer magnetizations. To this goal, we introduce a simple macrospin approximation with an effective Hamiltonian described by a bilinear form
\begin{equation}
	\bm{H}_\mathrm{MS} = -\Jeff\, \vec{M}_l(t) \cdot \vec{M}_r(t),
\end{equation}  
where each magnetic layer is characterized by a local magnetization $\vec{M}_{l,r}=\sum_{j}^{N_{S}}\vec{S}_j$, with $N_{S}=N_l=N_r$ being the number of spins in a layer. The validity of this approximate model is discussed in Appendix~\ref{sec:AppB}.
The time evolution of one macrospin described by this form is given by the equations of motion
\begin{equation}
	\frac{\partial \vec{M}_{\ell}}{\partial t}  
	=\Jeff\, \vec{M}_{\ell}\times\vec{M}_{\overline{\ell}},
\label{eq:5_MeanFieldEOM}
\end{equation}
where $\ell,\overline{\ell}\in\{l,r\}$ and $\ell\neq\overline{\ell}$. Under some simple assumptions (e.g., that $|\vec{M}_{\ell}|/N_S = 1$), these nonlinear coupled ordinary differential equations can be solved analytically by rotating the system to the plane of the limit cycle and then back. In accordance with the later investigated case of an open system we set the initial condition to a parallel formation of classical spins within a layer but perpendicular between the magnetic layers (as illustrated in Fig.~\ref{fig:SpinValveStructure}). In particular, initially $\vec{M}_l(t=0)$ points to $x$ direction and $\vec{M}_r(t=0)$ to $z$ direction. The initial condition of the electrons is set by exact diagonalization under the half-filling condition ($\overline{\mu}=0$). 
The solution of Eq.~\eqref{eq:5_MeanFieldEOM} with the above initial state reads (for details see Appendix~\ref{sec:AppB}): 
\begin{equation}
	\vec{M}_{\ell}(t) =
	\begin{pmatrix}
	 \cos\theta & 0 & \sin\theta\\
	 0 & 1 & 0\\
	 -\sin \theta & 0 & \cos\theta
	\end{pmatrix}
	\begin{pmatrix}
	\frac{N_S}{\sqrt{2}} \cos(\omega t+\phi_{\ell})\\
	\frac{N_S}{\sqrt{2}} \sin(\omega t+\phi_{\ell})\\
	N_S/\sqrt{2}
	\end{pmatrix}. \label{eq7:MeanField_Solution} 
\end{equation}
Here, the right vector represents the solution in the frame of the limit cycle and the left matrix is the reverse rotation around the $y$-axis to the original frame of the spin valve, where $\theta=\pi/4$, $\phi_l=0$ and $\phi_r=\pi$. 
The characteristic frequency is given by $\omega = \sqrt{2} N_S \Jeff$. To quantify the influence of the electronic spectrum on spin dynamics, we use a least-squares fitting of this analytical solution to the numerical data (obtained within the QC-EOM approach). This also allows us to test the validity of the macrospin approximation and, in some limiting cases, the precision of our numerical integration. Note that there are parameter regimes where we also need a second fitting parameter $\varphi$ which shifts the phases to $\phi_{l}=\varphi$ and $\phi_r=\pi+\varphi$ in cases where the limit cycle is reached only after some significant time.

\subsubsection{Spin valve dynamics}\label{sec:SpinValveDynamics}

\begin{figure}[!h]
\centering
\includegraphics[width=1.0\columnwidth]{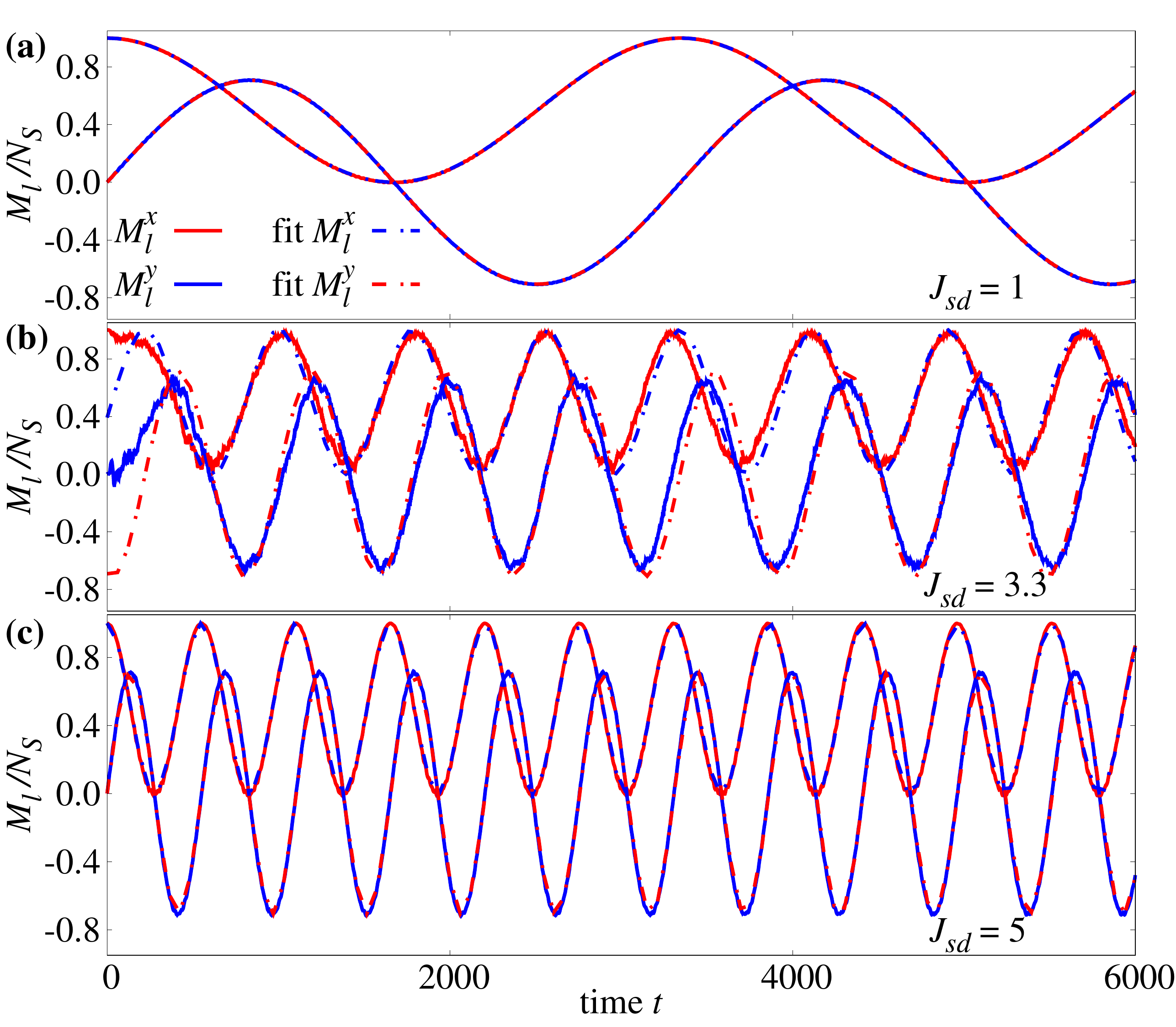}
\caption{Dynamics of the normalized layer magnetizations (solid lines) and their approximate analytical macrospin solution (dashed lines) with fitted $\omega$. The plotted data are for the first ferromagnetic layer calculated for a one dimensional spin valve of total length $N=20$ ($N_S\equiv N_{FM1}=N_{FM2}=5$) and couplings $\Jsd=1$ (a), $3.3$ (b) and $5$ (c).
\label{fig:NumMS}}
\end{figure}

We first demonstrate the validity of the macrospin approximation by comparing it with the numerical simulations. Fig.~\ref{fig:NumMS} depicts the dynamics of the magnetization in the first layer (solid lines) and its macrospin fit (dashed lines) for system size $N=20$ ($N_S=5$) and three $\Jsd$ values. The macrospin approximation fits the exact dynamics almost perfectly for $\Jsd=1$ because here the single-spin fluctuations are effectively suppressed already at small times. If we neglect the small fluctuations and oscillations imposed on top of the main dynamics, which are not visible on the scale presented in Figs.~\ref{fig:NumMS}(c), a similar conclusion can be drawn also for $\Jsd=5$. Interestingly, it is the case of intermediate coupling $\Jsd=3.3$ where the full dynamics becomes rather complicated, for example, it takes some transient time ($t\approx 500$) until the limit cycle is reached. Although the dominant precession frequency can be still extracted for this case, there is some modulation and the fit is far from perfect.  

\begin{figure}[!h]
	\centering
	\includegraphics[width=1.0\columnwidth]{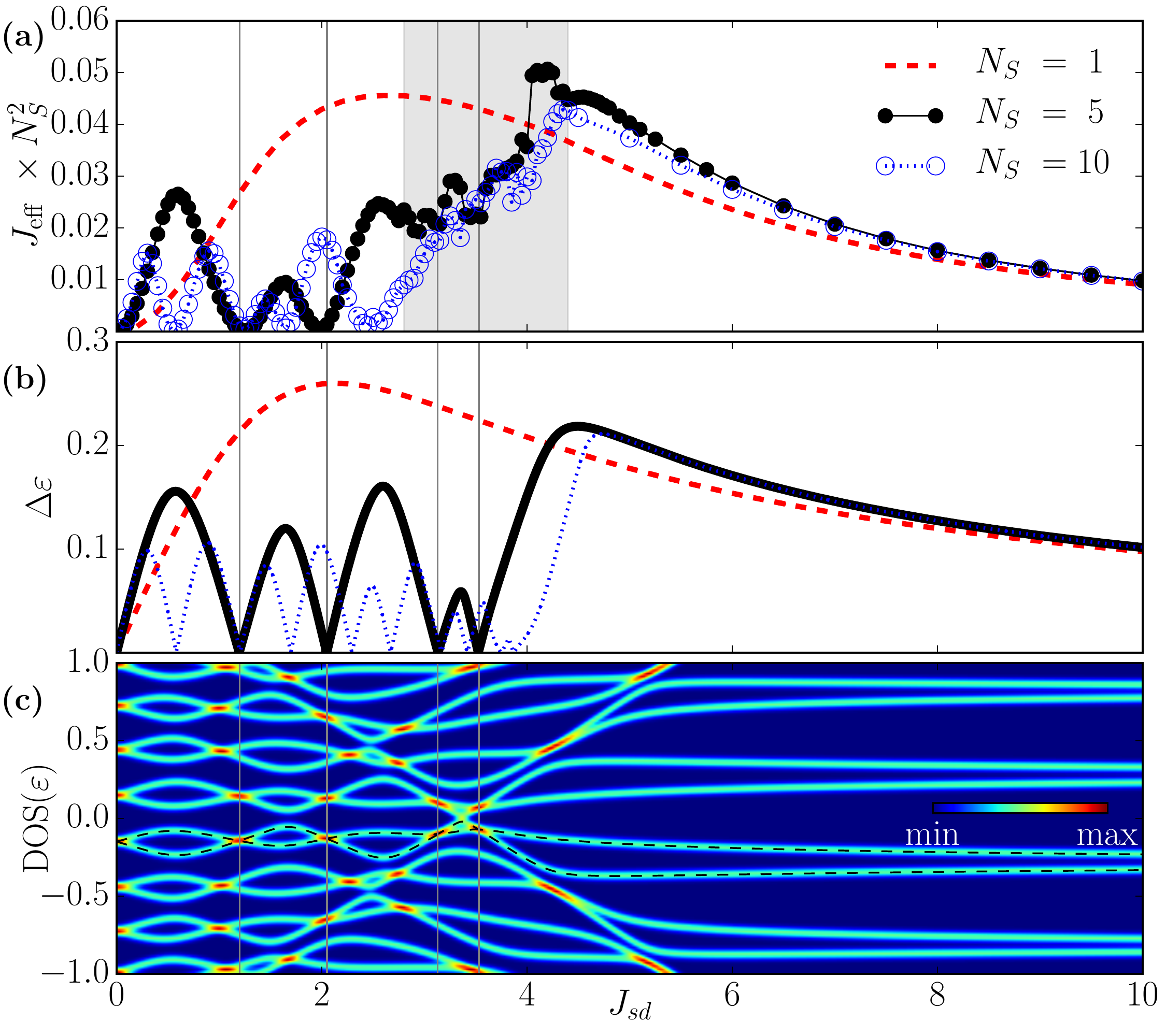}
	\caption{(a) Effective coupling $\Jeff$ (multiplied by $N^2_S$) extracted from least-square fit analysis of the numerical spin dynamics. The 
		shadowed area indicates a regime with high fitting uncertainty, i.e., where the analytical macrospin solution significantly departs from the full numerical one. (b) Energy difference  $\Delta \varepsilon$ between the two highest occupied energy levels [marked by dashed line in (c)] for $N_S=1,5$ and $10$. (c) Detail of the static density of states $\mathrm{DOS}(\varepsilon)$ for $N_S=5$. All presented $\Jsd$ dependencies were calculated for a one-dimensional chain with spacer layer size $N_\mathrm{NM}=12$ and exchange coupling $J_\mathrm{ex}=-1$.\label{fig:AnalysisPeaks}}
\end{figure}

To understand how the coupling $\Jsd$ affects the magnetization dynamics, we analyze a system with the spacer layer of length $N_\mathrm{NM}=10$ and three different sizes of magnetic layers. The $\Jsd$ dependence of the simplest $N_S=1$ case, plotted with the red dashed line in Fig.~\ref{fig:AnalysisPeaks}(a), shows a single broad maximum at $\Jsd\approx 2.5$. With increasing number of spins, the dependence becomes rather complicated. It exhibits several local maxima and minima for $N_S=5$ (black bullets) and $10$ (blue circles) and becomes monotonous only for $\Jsd\gtrsim 4.5$ where all $\Jeff \times N_S^2$ curves approach each other. 

This complicated behavior reflects the (static) electronic spectrum shown for the $N_S=5$ case in Fig.~\ref{fig:AnalysisPeaks}(c). Both the weak coupling $J_{sd}\lesssim 3$ and strong coupling $J_{sd}\gtrsim 4.5$ cases can be qualitatively understood by following the energy difference $\Delta \varepsilon$ [Fig.~\ref{fig:AnalysisPeaks}(b)] between the two highest occupied energy states in the static spectrum marked by the dashed line in Fig.~\ref{fig:AnalysisPeaks}(c).  Here, the energy difference $\Delta \varepsilon$ signalizes the magnitude of magnetic splitting (see the blue and red lines in Fig.~\ref{fig:SpecTriLayer}(a) for illustration). The effective coupling $\Jeff$ takes local minima when $\Delta \varepsilon$ approaches zero. The reason is that the nonmagnetic states do not couple to the classical spin and can not mediate the effective exchange coupling~\cite{smorka2021singlespin}. Consequently, the largest $\Jeff$ reflects the maximum in $\Delta \varepsilon$ and \emph{vice versa}. In the case of large spin-electron coupling $\Jsd$ the spectrum is divided into three bands and the states that are mostly localized to the ferromagnetic layers are far away from the Fermi level. Because we do not change the size of the spacer layer, the splittings $\Delta \varepsilon$ for various $N_S$ approach each other and the same pattern is followed by $\Jeff$.

The only regime where the fitted $\Jeff$ departs qualitatively from $\Delta \varepsilon$ (gray area in Fig.~\ref{fig:AnalysisPeaks}(a) for $N_S=5$) coincides with the splitting of the three bands illustrated in Fig.~\ref{fig:SpecTriLayer}. Here we observe the transition from metallic to insulating character of the valve (see also discussion in~\Cref{App:loc}).
This is accompanied by strong electron-induced spin fluctuations on a time scale much shorter than the main precession. These spin fluctuations lead to a deviation from the initial FM ordering which significantly modifies the electronic spectrum, and therefore also  $\Delta \varepsilon (t)$,  which can not be considered static anymore. This is clearly reflected in the $J_\text{eff}$ in this regime which does not follow the $\Delta \varepsilon(t=0)$ (for details see \Cref{App:SpinFluctuations}).
Nevertheless, we can conclude that the sensitivity of $\Jeff$ on the details of the electronic spectrum, in all above discussed regimes, underlines the importance of treating electrons as quantum particles instead of using effective classical approximations. 

\begin{figure}[!h]
	\centering
\includegraphics[width=1.0\columnwidth]{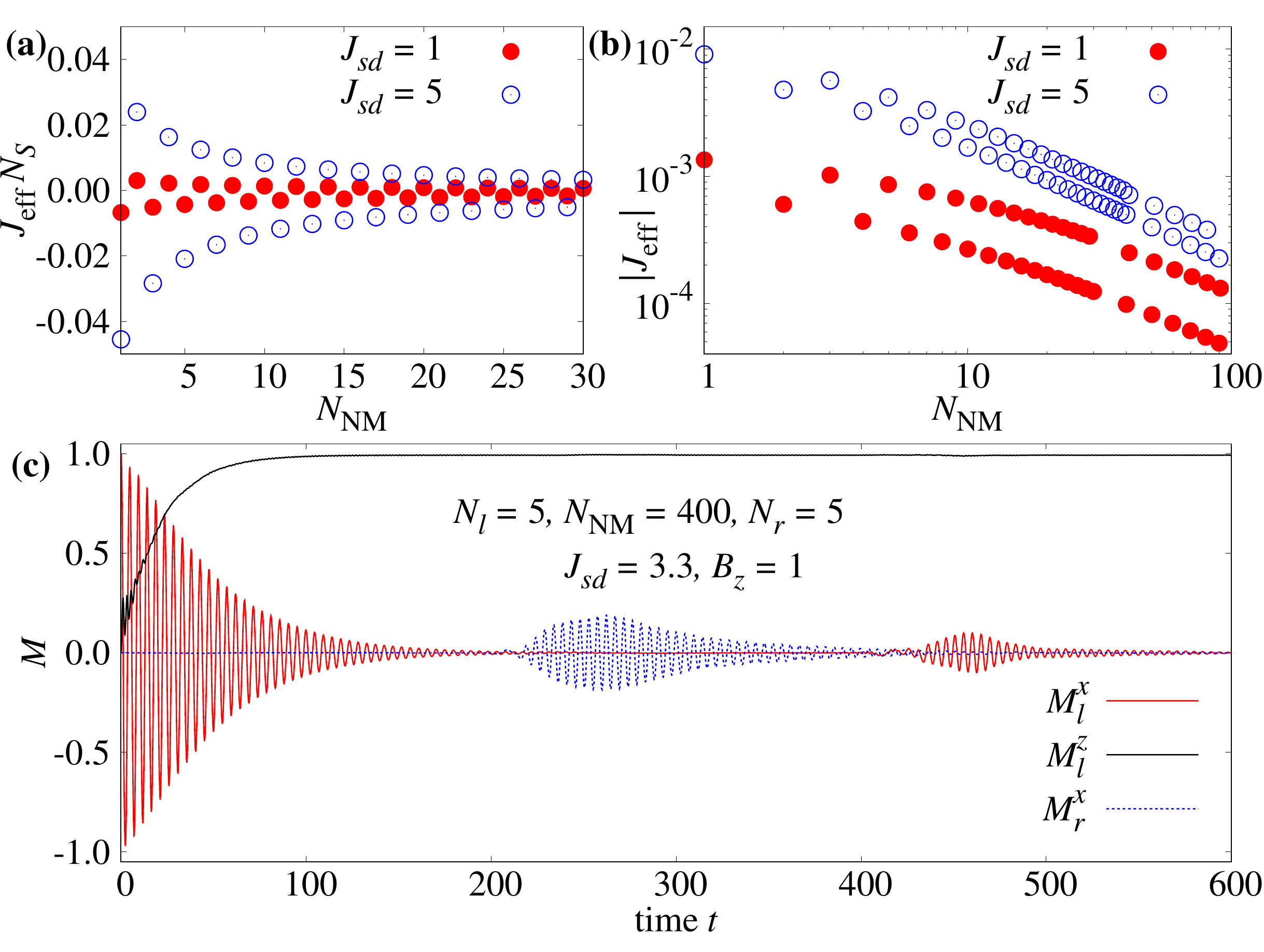}
	\caption{(a,b) Fitted effective coupling $\Jeff$, respective its magnitude $|\Jeff|$, as a function of the size of the non-magnetic spacer layer. (c) Normalized layer magnetization for left ($M^x_l$, $M^z_l$) and right magnetic layer ($M^x_r$) in spin valve with long non-magnetic layer $N_\mathrm{NM}=400$, $\Jsd=3.3$ in homogeneous magnetic field $B_z=1$.\label{fig:tdL400}}
\end{figure}

Outside the regime $3\lesssim J_{sd}\lesssim4.5$, the macrospin approximation works well also in the case of varying width of the spacer layer. Figs.~\ref{fig:tdL400}(a,b) show the dependencies of $\Jeff$ on $N_\mathrm{NM}$ for weak $\Jsd=1$ and strong spin-electron coupling $\Jsd=5$. The alternation of $\Jeff$ between ferromagnetic and antiferromagnetic character (a) as well as the algebraic decay with increasing $N_\mathrm{NM}$ (b) are in qualitative compliance with previous results~\cite{Parkin1990,Bruno1991,litvinov1998RKKY}. These features are captured already by perturbation approaches, e.g., the theory of Ruderman-Kittel-Kasuya-Yosida (RKKY) interaction~\cite{Saremi2007,Bunder2009,Rusin2020} which for a one dimensional electron gas predicts $J_\textrm{eff}\propto J_{sd}^2[\mathrm{Si}(\pi N_\mathrm{NM})-\pi/2]$ where $\mathrm{Si}(x)$ is the sine integral function~\cite{Rusin2020}. The large difference in magnitude between the odd and even $N_\mathrm{NM}$ shown in Figs.~\ref{fig:tdL400}(b), results from the difference between the polarizations of states at the Fermi level for chains of odd and even point numbers.

Although useful, the above fitting to the macrospin dynamics is bound to fail in a more realistic setup. The reason is that the above mean-field theory cannot capture some important features of the whole dynamics. For example, it actually takes a finite time for electrons to react to a new position of the classical spins~\cite{Sayad_2015,Sayad_2016_epl,Sayad_2016} and carry the excitation from one magnetic layer to the other. For a long spacer layer, this can lead to a significant delay between the dynamics of the two magnetic layers. In addition, the microscopic dynamics of electrons generates a time-retarded damping in the dynamics of classical spins~\cite{Sayad_2015,Petrovic18}. We illustrate this in Fig.~\ref{fig:tdL400}(c) using a long nonmagnetic layer $N_\mathrm{NM}=400$ with the same initial condition as before, however, we also introduce an external magnetic field in the $z$ direction, $B_z=1$, which triggers Larmor oscillations in the first magnetic layer. It is clear that it takes a finite time ($t \approx N_\mathrm{NM}/2\gamma$) before the excitation from the first magnetic layer (red curve) reaches the second one (blue curve). What is even more important in the context of our work is that the presence of a long spacer layer leads to a relaxation of the spin oscillations. 

A similar relaxation effect can be obtained even for a short spacer layer by coupling the spin valve to semiinfinite metallic leads~\cite{Elbracht2020leads,smorka2021singlespin}. 
In addition, coupling to the leads allows us to address a system influenced by an external voltage drop~\cite{smorka2021singlespin}.  

\subsection{Voltage driven spin valve}\label{sec:Driven}

To investigate the influence of nonequlibrium charge and spin currents, resulting as a consequence of an external voltage drop, on the magnetization dynamics, we now turn our attention to a two-dimensional spin valve ($N=12\times 4$, $N_S=4\times4$) with non-collinear magnetization sandwiched between two semi-infinite metallic leads (Fig.~\ref{fig:SpinValveStructure}). As before, we set the model parameters with the aim to make the analysis tractable by simplifying the spin dynamics. 
We set an intermediate coupling $\Gamma\equiv\Gamma_l=\Gamma_r=1$ and use a fixed temperature in the leads $T=0.025$. The intermediate $\Gamma$ reduces reflection of the conducting electrons at the valve-lead interface~\cite{smorka2021singlespin}  (typical for weak $\Gamma$), provides sufficient broadening~\cite{Zonda2019} but does not dominate over other energy scales of the model. In addition, because $\Gamma \gg T$, finite temperature effects are suppressed and, therefore, not discussed in detail here. The exchange coupling is set to $J_\mathrm{ex}=-1$ which is strong enough to allow us to represent the dynamics of a magnetic layer by its normalized net magnetization $\bm{M}_\ell/N_S$ (macrospins)~\footnote{The size of the resulting normalized macrospin is not fixed to one, however, for the chosen parameters it is very close to one especially at longer times.}. The anisotropy in the left layer is set to be large, $K_l=0.4$ (therefore pinned layer PL) and points to $x$-direction. The anisotropy in the second ferromagnetic layer is set to $K_r=0.02$ (therefore free layer FL) and points to $z$-direction. 

\begin{figure}[!h]
	\centering
	\includegraphics[width=1.0\columnwidth]{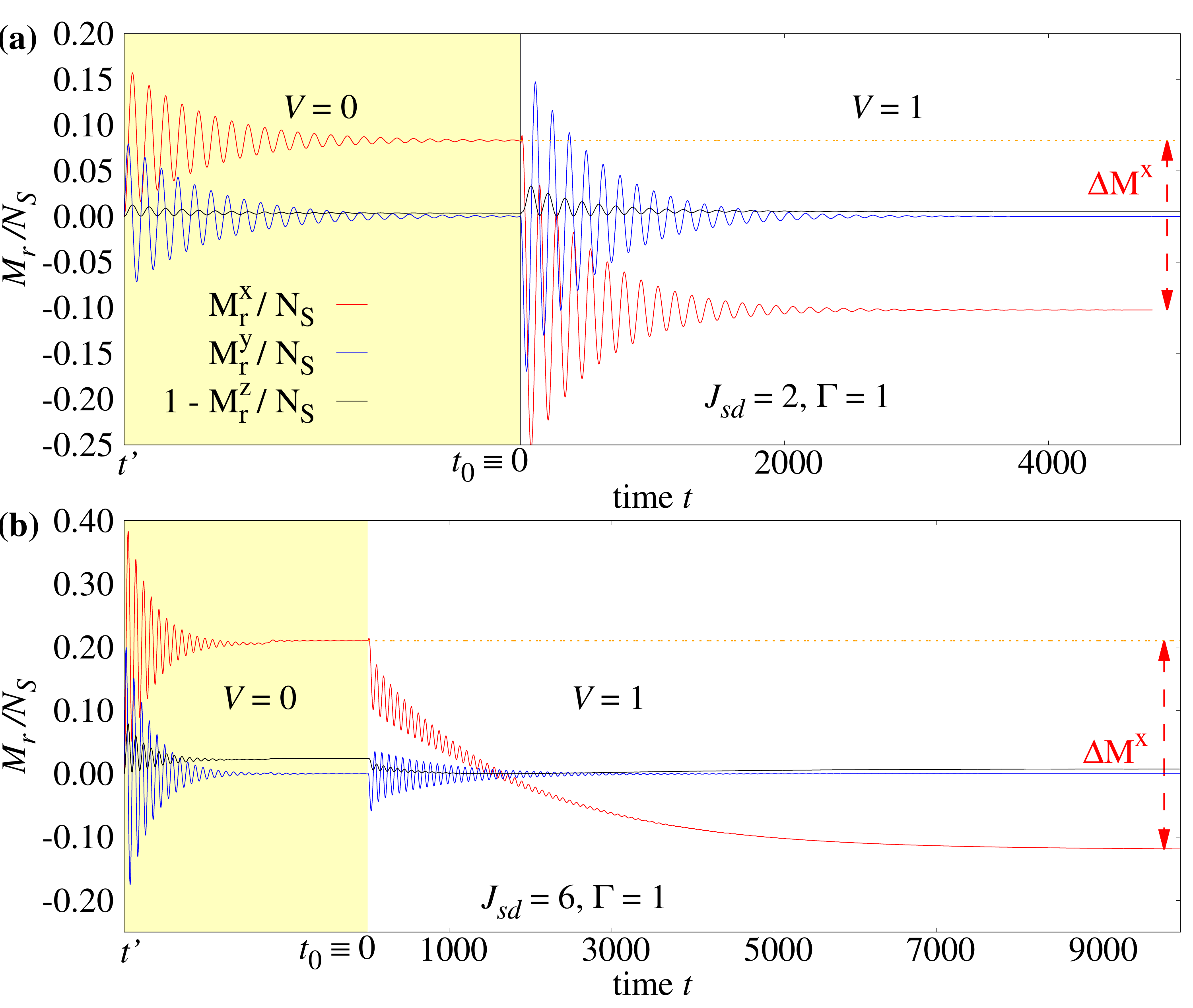}
	\caption{Examples of dynamics of normalized magnetization in the free layer calculated for two-dimensional spin valve with size $N=12\times 4$, $N_S=N_\mathrm{PL}=N_\mathrm{FL}=4\times 4$ and model parameters: $\Jsd=1$ (a), $\Jsd=5$ (b), and $\Gamma=1$, $K_l=0.4$, $K_r=0.02$. Yellow background marks stage one with $V=0$. At $t=t'\{\equiv0\}$ a quench to finite symmetric voltage drop $V=1$ is introduced. \label{fig:Mag_td}}
\end{figure}

We work within a partition-free approach~\cite{zhang2013first,ridley2018formal} where the system-reservoir coupling $\Gamma$ is assumed to be finite at all times. To investigate the effect of a finite bias voltage on the magnetization dynamics, we employ a three-stage switching protocol 
which leads to the net layer magnetization dynamics illustrated in Fig.~\ref{fig:Mag_td}.

\emph{Stage 0:} We assume that at $t < t'$, the classical spins in each layer are perfectly parallel to the direction of the layer anisotropy: ${M}^x_\mathrm{PL}=N_S$ and $M^z_\mathrm{FL}=N_S$. The spin valve is in equilibrium with the electronic reservoirs with $V=0$. This is ensured by solving Eq.~\eqref{eq:EOM3} and Eq.~\eqref{eq:EOM_AuxCur} for $\frac{\dif}{\dif t}\rho = 0$ and $\frac{\dif}{\dif t} \Pi_{\ell,p}(t) = 0$. \emph{Stage 1:} At $t=t' \ll 0$ we ease the condition of perfect alignment of the spins with the anisotropy fields. Therefore, the effective coupling between the magnetic layers, alike the one discussed for the isolated spin valve, triggers spin dynamics. 
Because of the damping driven mainly by the dissipation of polarized electrons, the subsystem of localized spins relaxes towards a new (static) configuration (this stage is marked by the yellow background in Fig.~\ref{fig:Mag_td}). 
\emph{Stage 2:} At $t=t_0=0$, when the system has already relaxed into the equilibrium state, we induce again a nonequilibrium situation by suddenly switching on a finite bias voltage $V\neq 0$. However, as we argue below, the way how the drop is introduced plays a crucial role in the transient dynamics as well as in the steady-state. Therefore, we introduce the voltage in two distinct ways. Either by shifting the chemical potential of both leads around the equilibrium state ($\mu_l=\mu_r=\overline{\mu}=0$): $V=\mu_l-\mu_r$ with $\mu_l=-\mu_r$ which we call the symmetric case (in the sense of $|\mu_l|=|\mu_r|$), or by moving only the chemical potential of the left reservoir coupled to the pinned layer: $V'=\mu_l$ with $\mu_r=\overline{\mu}=0$ addressed in the text as the asymmetric case. Note that these two ways mimic different physical realizations. For example in the case of small system the symmetric voltage drop can model a gate-tunable junction or a bridge and the asymmetric one a scanning tunneling microscope (STM) like geometry where the chemical potential of the surface electrode is aligned with the gate induced electrochemical potential.

In our model the difference between these two scenarios lies in the position of the chemical potential of the right lead with respect to the equilibrium electrochemical potential $\overline{\mu}$ fixed by an auxiliary gate. This difference is important for understanding the results. Basically, the drop of the chemical potential in the leads changes the electron occupation in the valve, however, the non-equilibrium distribution of the charge does not modify $\overline{\mu}$ fixed by the auxiliary gate and, therefore, does not shift the spectrum of the valve. A self-consistent calculation adjusting $\overline{\mu}$ after introducing the leads, which might be necessary for systems without a gate, is not part of the presented model calculations. A detailed recipe on how to address the problem of gauge-invariant density matrix in steady state nonequilibrium linear response calculations for systems without a gate can be found in Ref.~\cite{Mahfouzi2013}. We show in~\cref{App:AppLR} that for the method used here the two discussed voltage-drop scenarios, although different in general, lead to the same charge current in the linear response regime.     

In practice, we calculate the stages zero and one only once for all required voltage drops for the same system parameters. At $t=t_0\equiv0$ we then store the state of the system, meaning the orientation of localized spins, the equilibrium single particle density matrix $\rho_\mathrm{eq}(t=0)$ and all auxiliary current matrices, and use it as the initial condition for the second stage.    
This gives us a well-defined \emph{initial} equilibrium state for the coupled system at $V=0$ which differs from the stage zero result. This is crucial, not only because it enables us to calculate the actual change of measured quantities due to the finite voltage, e.g., the magnetization change $\Delta \vec{M}(t)=\vec{M}(t)-\vec{M}_\mathrm{eq}$ or current-driven torques, but it also allows us to investigate a well defined relaxation.

\subsubsection{Transient dynamics}
There are various relevant time scales associated with the dynamics of the coupled system. Some are related to two distinct anisotropies and lead to different spin oscillations in the pinned and free layer. Others are demonstrated already by the examples of time evolution of the magnetization in the free layer shown in Fig.~\ref{fig:Mag_td}. 
The difference between the time at which the system reaches the steady-state magnetization and the relaxation time of the magnetization oscillations depends on the parameters of the model and the voltage drop. 
For example, for $\Jsd=2$ and $V=1$ in Fig.~\ref{fig:Mag_td}(a) the oscillations are centered around the steady state values of $M_r^{x,z}$ quite early on ($t\sim 500$), but the relaxation time of the oscillations themself is much longer. On the other hand, in the case $\Jsd=6$ and $V=1$ magnetization $M_r^x$ decreases to its steady state value only very slowly and it is here the longer of the two mentioned timescales. Nevertheless, these are not the only relevant time scales of the transient dynamics.
\begin{figure}[!ht]
	\centering
	\includegraphics[width=1.0\columnwidth]{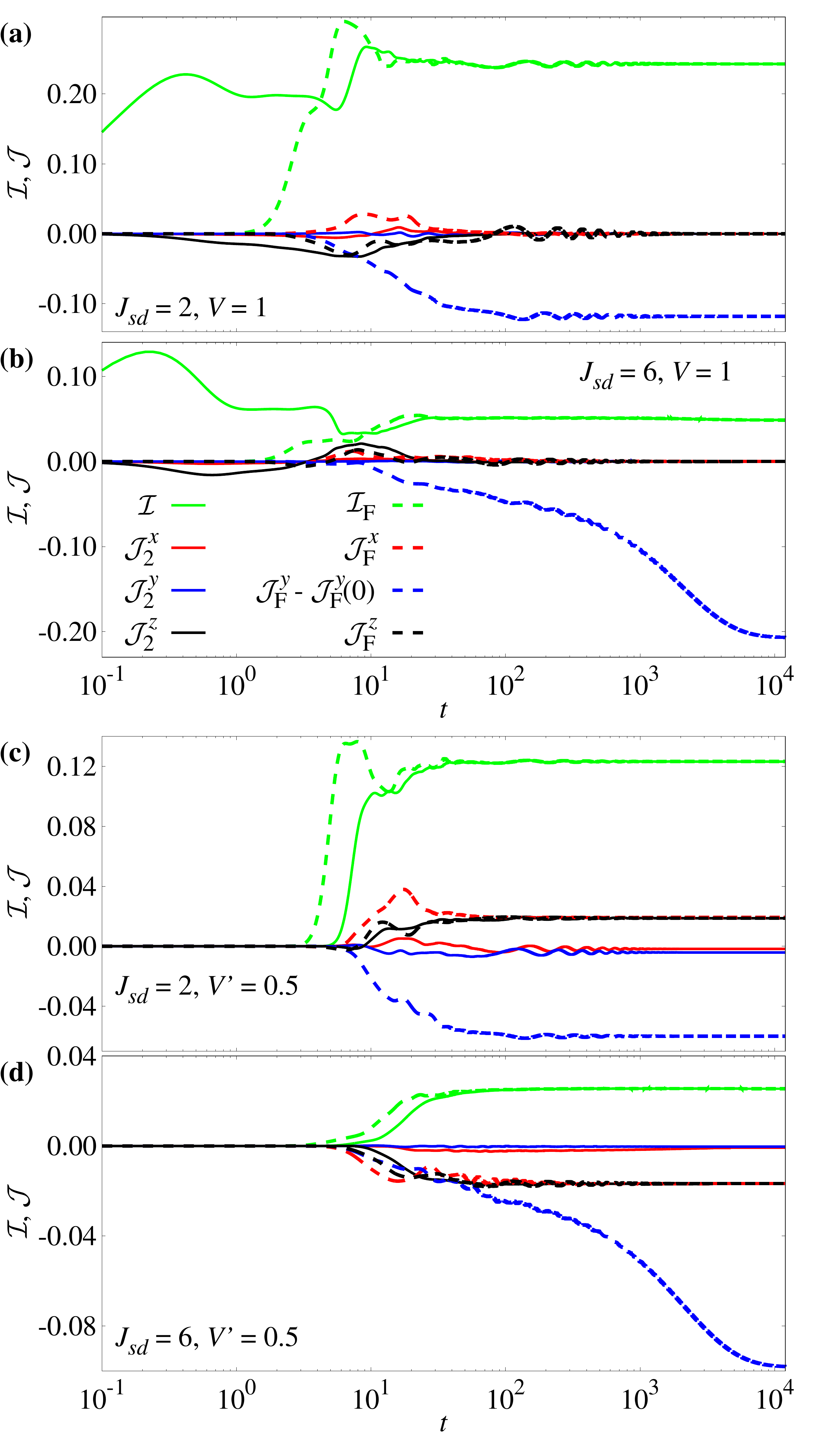}
	\caption{Second stage of the time evolution of charge (${\cal I}$) and spin currents (${\cal J}$) calculated for symmetric voltage drop $V=1$ with $\Jsd=2$ (a) and $\Jsd=6$ (b) and asymmetric voltage drop $V'=0.5$ with $\Jsd=2$ (c) and $\Jsd=6$ (d). The solid lines show currents measured at the FL-lead interface and dashed lines represent respective currents at SL-FL interface. The color key is the same for all panels.\label{fig:Ispin_td}}
\end{figure} 

Fig.~\ref{fig:Ispin_td} shows examples of the time evolution of charge and spin currents in the second stage for the same model parameters as in Fig.~\ref{fig:Mag_td}. The voltage in Fig.~\ref{fig:Ispin_td}(a) and (b) was introduced by a symmetric voltage drop ($V=1$) and in panels (c) and (d) by an asymmetric one ($V'=0.5$). The solid lines represent currents at the FL-lead interface, namely, green for the total charge current [Eq.~\eqref{eq:Cur}], red, blue, and black for the $x,y,z$-component of the spin current, respectively [Eq.~\eqref{eq:SCur}]. 
The dashed lines of the same colors are local currents measured at the interface of the spacer and free layer [Eqs.~\eqref{eq:Curl},~\eqref{eq:SCurl}].

Almost all currents are zero or negligible at $t=0$ (not shown because of the logarithmic scale). The only exception is the equilibrium local spin current ${\cal J}^y_F(t=0)\neq0$. This is related to the effective exchange interaction between PL and FL and torque resulting from it that tilts the spins of the magnetic layers away from the direction of the anisotropy even in equilibrium (see yellow stage in Fig.~\ref{fig:Mag_td}). In Fig.~\ref{fig:Ispin_td} we subtracted this equilibrium component from the nonequilibrium value $\overline{{\cal J}}^y_F(t)={\cal J}^y_F(t)-{\cal J}^y_F(0)$ because here we want to investigate the influence of the finite bias voltage. In accordance with other components, we address this difference simply as the local current for brevity.

Application of a symmetric bias voltage induces transient currents simultaneously through both system-lead interfaces.
Therefore, as shown in Fig.~\ref{fig:Ispin_td} (a) and (b), there are significant currents flowing through the right system-lead interface already after a very short time. 
Approximately at $t\approx 2$ the excitation from the right spin valve interface arrives to the SL-FL interface leading to the formation of local currents there. Next, at $t\approx 4$ the excitation from the left system edge reaches the SL-FL interface, which marks sudden changes in the profile of the currents. 
Considering longer times, because the voltage drop is symmetric, the spin currents at the system-lead interface vanish in the steady state. Qualitatively this can be understood following the profile of the equilibrium transmission function polarization (see the discussion in ~\cref{sec:AppC}) which is antisymmetric around the Fermi level. Therefore, the relevant Fermi window contains compensating spin-resolved transmission channels and the steady-state spin current vanishes. 

The only non-negligible long-time nonequilibrium spin current is $\overline{{\cal J}}^y_F$, which is therefore the sole component of the current-driven torque in Eq.~\eqref{eq:torque}. This means that the spin currents are fully absorbed by the magnetic layers.  As we will discuss in the next section, this has important consequences for the relaxation and for the steady state magnetization.  It also means that the difference in magnetization ($\Delta \bm{M}_{l,r}$) is driven primarily by the effective exchange interaction between the magnetic layers. However, contrary to the isolated system discussed in~\cref{sec:isosv}, this effective coupling might be strongly affected by the electronic states which are far away from the Fermi level of the spin valve. In addition, the nonequilibrium system density of states depends on the voltage drop and the orientation of the spins and is therefore time dependent. The slow change of $\overline{{\cal J}}^y_F(t)$ in  Fig.~\ref{fig:Ispin_td}(b) for $\Jsd=6$ can be attributed to the time evolution of DOSh, which is expected to change more for the strongly interacting case ($\Jsd=6$) than weakly coupled electrons and localized spins ($\Jsd=2$). 
     
The asymmetric voltage-drop examples shown in Figs.~\ref{fig:Ispin_td} (c) and (d) differ from the symmetric case. Because the voltage drop is introduced only at the side of the pinned layer, the free layer stays for a while in equilibrium with the right lead. Finite local currents appear around $t\approx 4$ when the excitation from the left edge of the spin valve reaches the SL-FL interface and only later we observe finite currents at the right system interface. Because the voltage drop probes the spin-dependent transmission function asymmetrically, the spin currents at negative and positive energies do not compensate each other and saturate to finite values.  Consequently, there are finite current-driven torques in both $x$ and $y$ directions even at long times.

\subsubsection{Relaxation}

\begin{figure}[!ht]
	\centering
	\includegraphics[width=1.0\columnwidth]{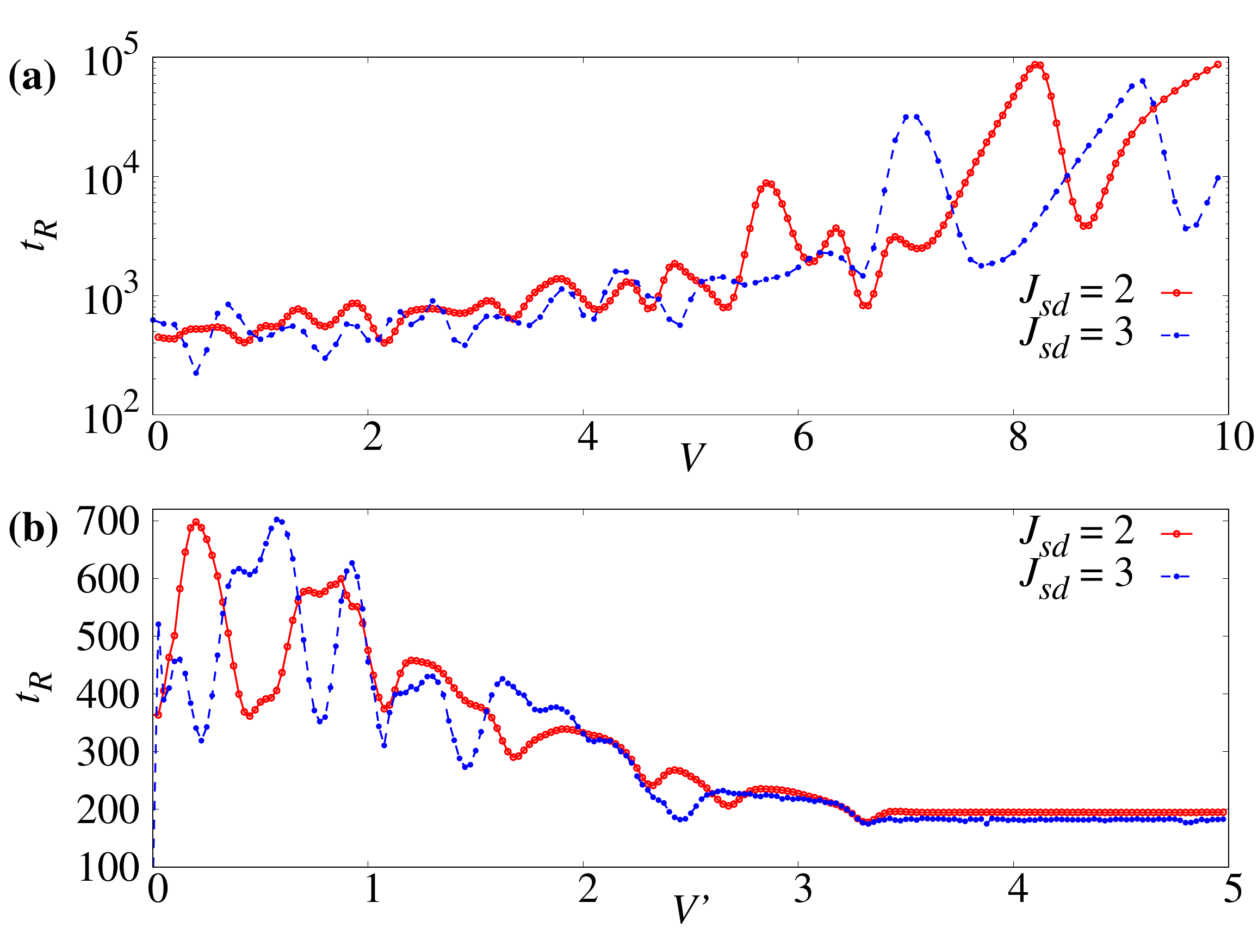}
	\caption{Relaxation time estimated from the decay of the oscillations of the $x$-component in the FL magnetization. Panel (a) shows the symmetric voltage case (note the logarithmic $y$-scale), panel (b) the asymmetric one.  \label{fig:retime}}
\end{figure}

The quantity that clearly demonstrates the qualitative difference between the symmetric and asymmetric voltage drop is the relaxation time $t_R$ of the magnetization oscillations (Fig.~\ref{fig:retime}). 
We estimate $t_R$ by fitting the envelope of $M_r^x$ oscillations in the second stage of the evolution by the exponential formula
\begin{equation}
 E_M(t)=A \exp [-t/t_R]+M_r^x(t\rightarrow\infty),
 \label{eq:fitformula}
 \end{equation} 
where the amplitude $A$ and the relaxation time $t_R$ are the fitting parameters and $M_r^x(t\rightarrow\infty)$ is the extrapolated steady-state magnetization component. We disregard in the fitting procedure the initial evolution in the second stage ( typically up to $t\approx 100-300$) to avoid the distortions from the complicated short-time dynamics discussed above. We focus on the weakly coupled cases $\Jsd=2$ and $3$ to avoid the long time scales typical for strong $sd$ coupling. 
 
The fitted relaxation time $t_R$ shows a qualitatively different dependence on voltage for the symmetric [Fig.~\ref{fig:retime}(a)] and asymmetric [Fig.~\ref{fig:retime}(b)] cases.
The relaxation time calculated for the symmetric case is changing by several orders of magnitude with increasing voltage $V$. It grows to very large values $t_R \approx 10^5$ at high voltages (see also the discussion on the numerical precision in ~\cref{sec:AppA}).  
On the other hand, the relaxation time in Fig.~\ref{fig:retime}(b) is relatively stable. It changes within five hundred time units and saturates for high voltage $V'$, where
$t_R$ is several orders of magnitude smaller than for the symmetric case. 
\begin{figure}[!ht]
	\centering
	\includegraphics[width=1.0\columnwidth]{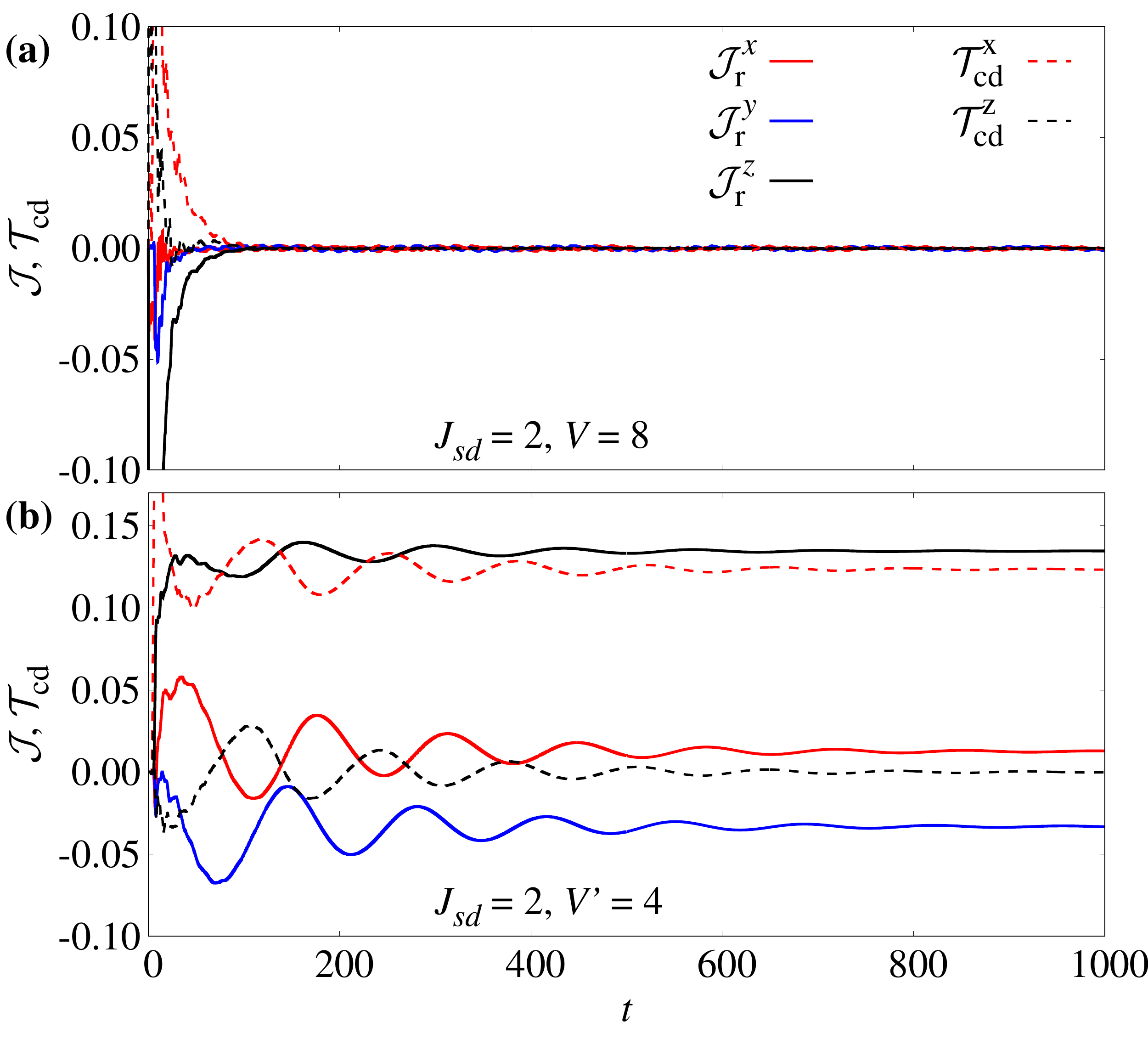}
	\caption{Detail of the time evolution of spin currents (${\cal J}$) and current-driven torques (${\cal T}_\mathrm{cd}$) calculated for symmetric voltage drop $V=8$ with $\Jsd=2$ (a) and asymmetric voltage drop $V'=4$ with $\Jsd=2$ (b)\label{fig:Ispin2_td}.}
\end{figure} 

\begin{figure}[!ht]
	\centering
	\includegraphics[width=1.0\columnwidth]{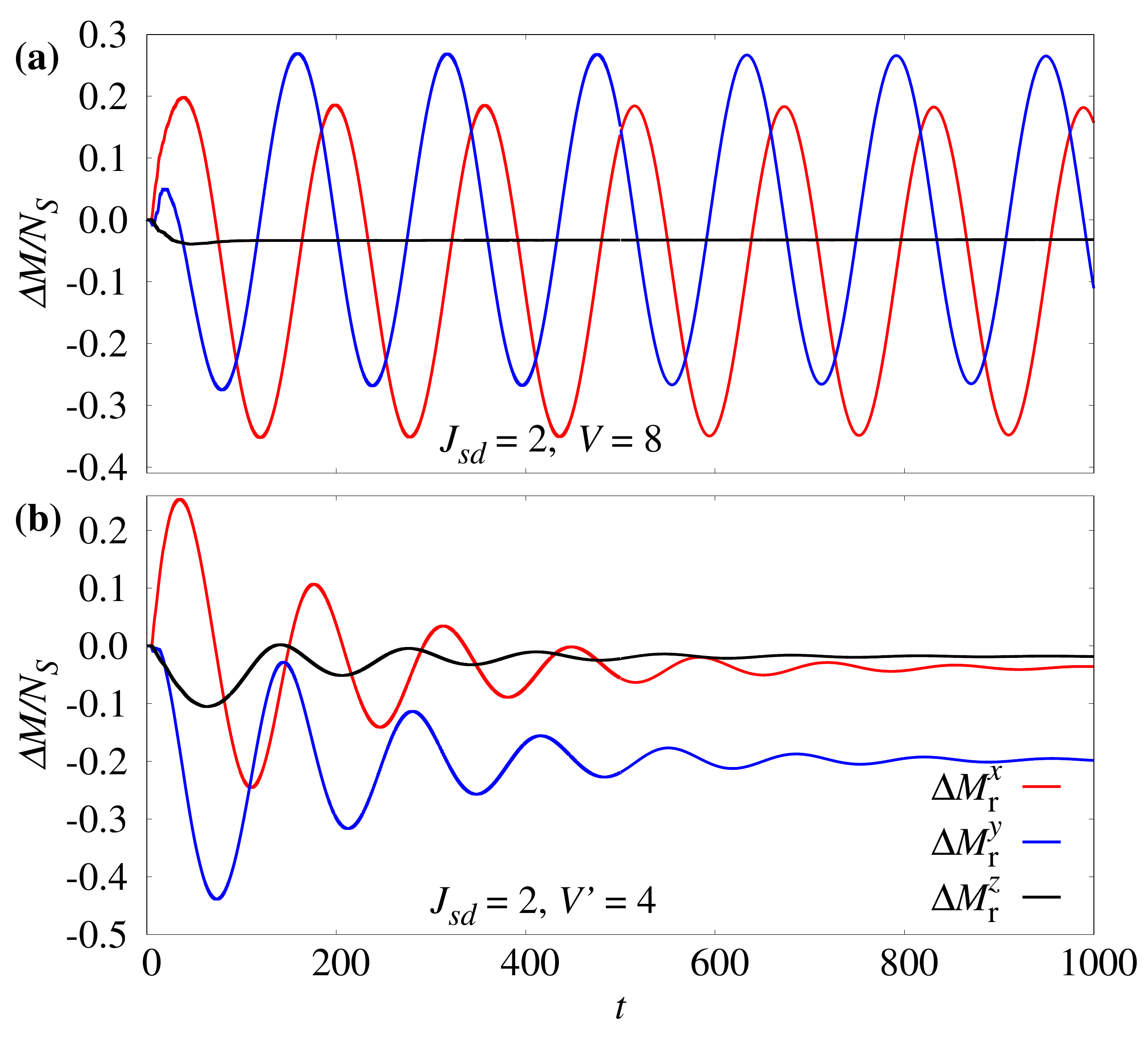}
	\caption{Detail of the time evolution of non-equilibrium magnetization difference ($\Delta\bm{M}/N_S$) calculated for symmetric voltage drop $V=8$ with $\Jsd=2$ (a) and asymmetric voltage drop $V'=4$ with $\Jsd=2$ (b)\label{fig:Mag2_td}.}
\end{figure}

This significant discrepancy in relaxation time can be attributed to the differences in the spin currents at the system-lead interface and related torques. The damping of the localized spin dynamics comes from the interaction with the leads, which act as reservoirs that carry away spin excitations from the system. However, this is possible only when they couple to the spin-resolved electronic states, i.e., when there are significant spin currents flowing between the system and the leads. 
This is not the case for the symmetric voltage drop as it is illustrated in Fig.~\ref{fig:Ispin2_td}(a) where we show the charge currents and current-driven torques for $V=8$ and $\Jsd=2$. Both currents and relevant torques are quickly diminishing, and hence do not exert any significant torque on the localized spins. Therefore, the magnetization shows a Larmor-like precession due to the effective fields as illustrated in  Fig.~\ref{fig:Mag2_td}(a). All this is in clear contrast with the respective asymmetric case ($V'=4$) illustrated in Fig.~\ref{fig:Ispin2_td}(b) and Fig.~\ref{fig:Mag2_td}(b).

\begin{figure}[!ht]
	\centering
	\includegraphics[width=1.0\columnwidth]{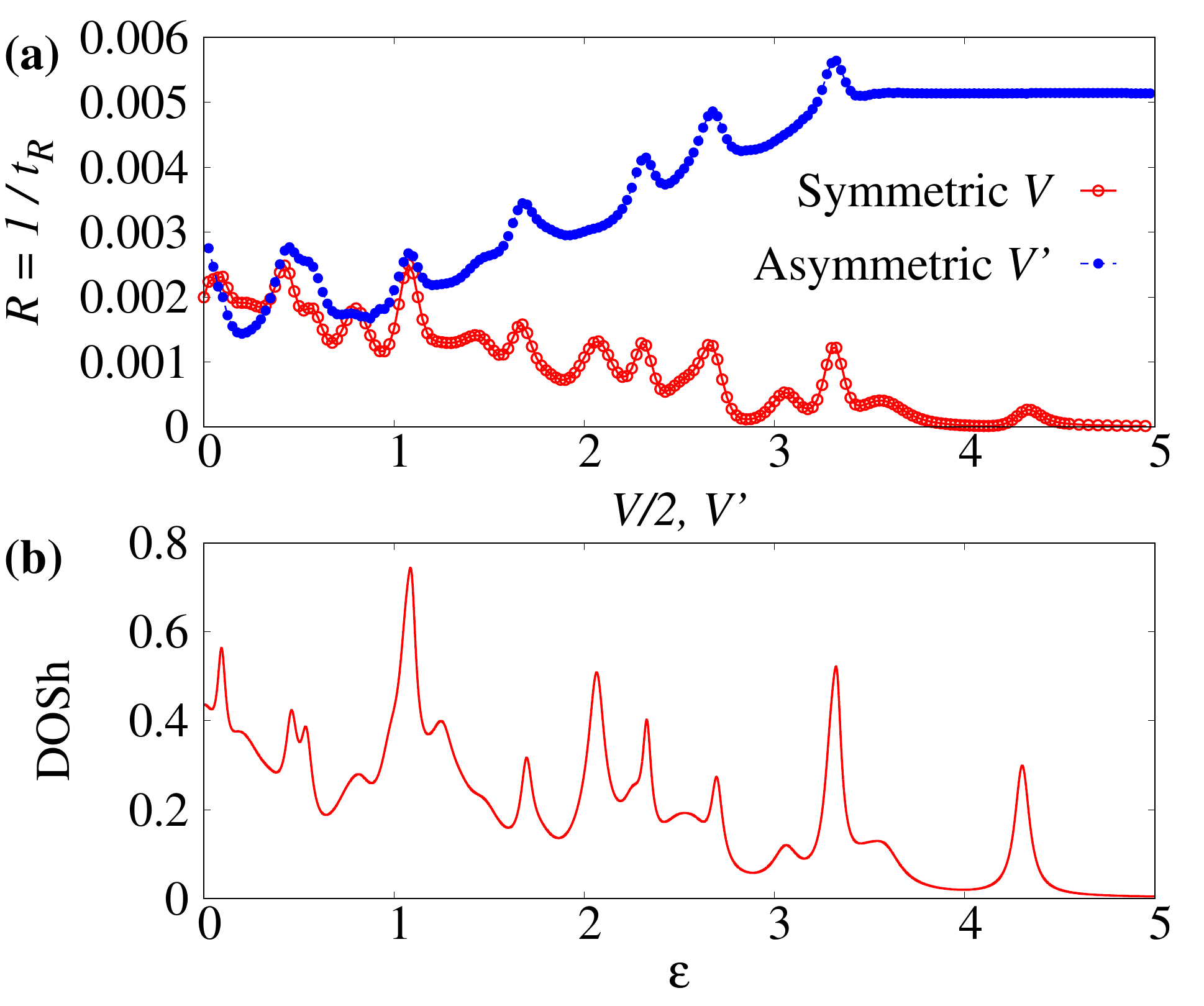}
	\caption{(a) Relaxation rates calculated for symmetric (red) and asymmetric (blue) voltage drop at $\Jsd=2$. The $x$-scale for the symmetric voltage is scaled by factor $0.5$ to align voltages with the same chemical potential $\mu_l$ probing energies $\varepsilon$. 
    (b) Equilibrium density of states of the heterostructure. \label{fig:rrAS_td}}
\end{figure}

However, the dependence of the relaxation time on the voltage drop is far from monotonous. The origin of the complicated profile can be traced to the density of states. In Fig.~\ref{fig:rrAS_td}(a) we show the relaxation rates $R=1/t_R$ at $\Jsd=2$ calculated for symmetric (red) and asymmetric (blue) voltage drop. The sharp maxima in both $R$ curves follow the profile of DOSh in Fig.~\ref{fig:rrAS_td}(b) calculated for the equilibrium spin configuration at $V=0$. This can be attributed to the boost of relaxation whenever the chemical potential of the leads is aligned with the (polarized) states in the system. Such a boost is in compliance with single spin studies~\cite{nunez2008effective,smorka2021singlespin,filipovic2013spin,hammar2016time} and shows the importance of the correct treatment of the electronic spectrum. Interestingly, for the symmetric voltage drop, the boost happens even for states with energies close to the edge of the spectrum. These are localized predominately on the ferromagnetic layers and as such practically do not contribute to the steady state transport (see discussion in \cref{sec:AppC}). However, these states can contribute to the relaxation if they are aligned with the chemical potential of the neighboring lead. In the asymmetric case, this contribution is overshadowed by the damping regulated by the finite spin currents.  

\subsubsection{Steady state}

\begin{figure}[h!]
	\centering
	\includegraphics[width=1.0\columnwidth]{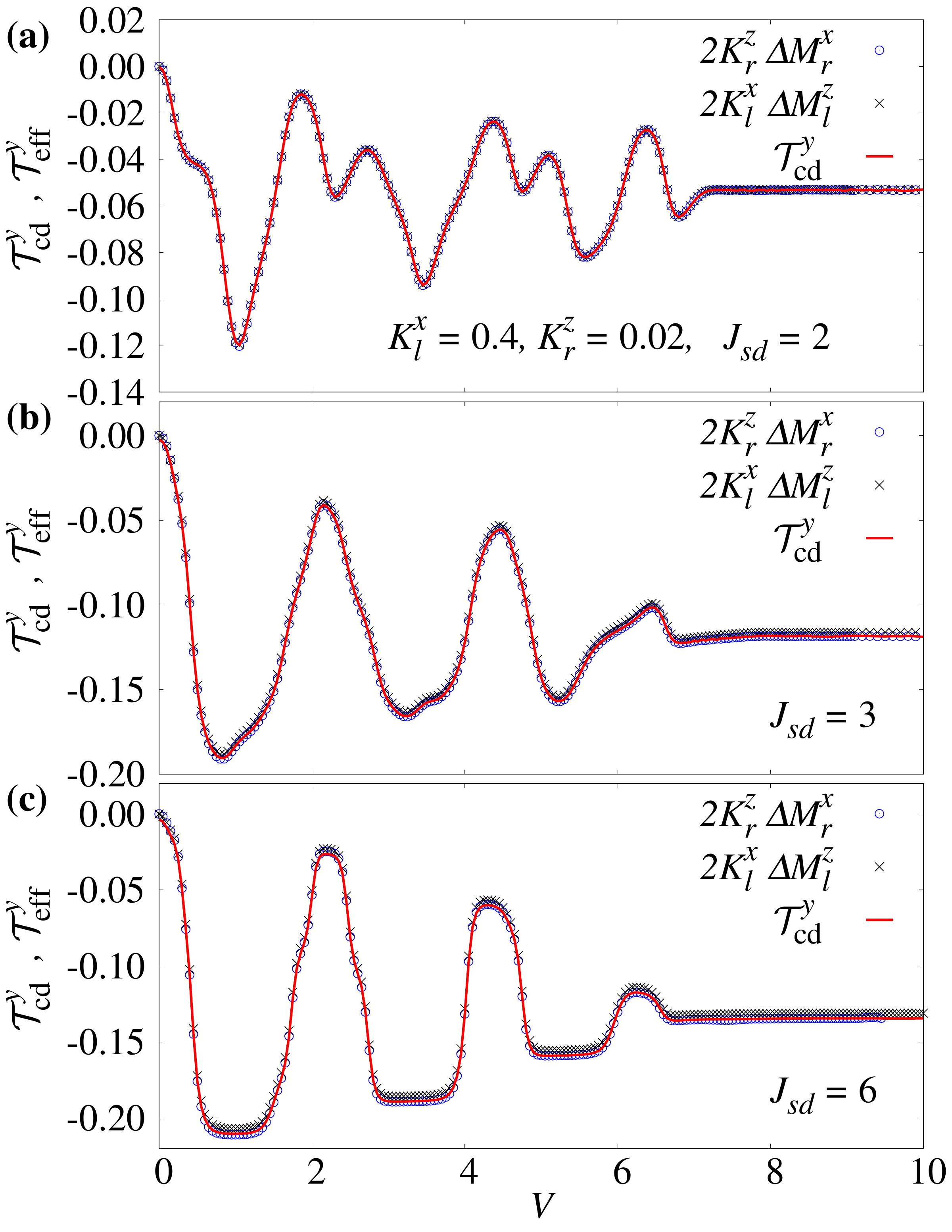}
	\caption{Comparison of the steady state current-driven torques ${\cal T}_\mathrm{cd}$ and effective local torques 
	$\Teff=2 K_\ell\bm{e}_\ell \times \Delta \bm{M}$ 
	calculated as function of $V$ for symmetric voltage drop. \label{fig:ststLR}}
\end{figure}

Fig.~\ref{fig:ststLR} shows (up to a constant factor) the steady state magnetization difference $\Delta \bm{M}$ as a function of the symmetric voltage drop (circles and crosses in the figures). The dependence is rather complicated and does not straightforwardly follow the equilibrium DOSh even for small $\Jsd$ [compare Fig.~\ref{fig:ststLR}(a) and  Fig.~\ref{fig:rrAS_td}(b)]. Nevertheless, it can be fully explained by considering the current-driven torques acting on the magnetic layers.
  
In the steady state, the net sum of all torques in the system is zero. In our case, the dominant contribution to the effective local fields acting on the localized spin, and counteracting the current-driven torques, should come from the misalignment of the spin with the anisotropy field and from the interaction with neighboring localized spins through $J_\mathrm{ex}$ [see Eq.~\eqref{eq:EOM2}]. However, assuming that the classical spins are aligned with each other, we can approximate the net effective torque acting on the right magnetic layer by 
$\Teff \approx 2 (K_r\bm{e}_r) \times \Delta \bm{M}_r$.
\begin{figure}[h!]
	\centering
	\includegraphics[width=1.0\columnwidth]{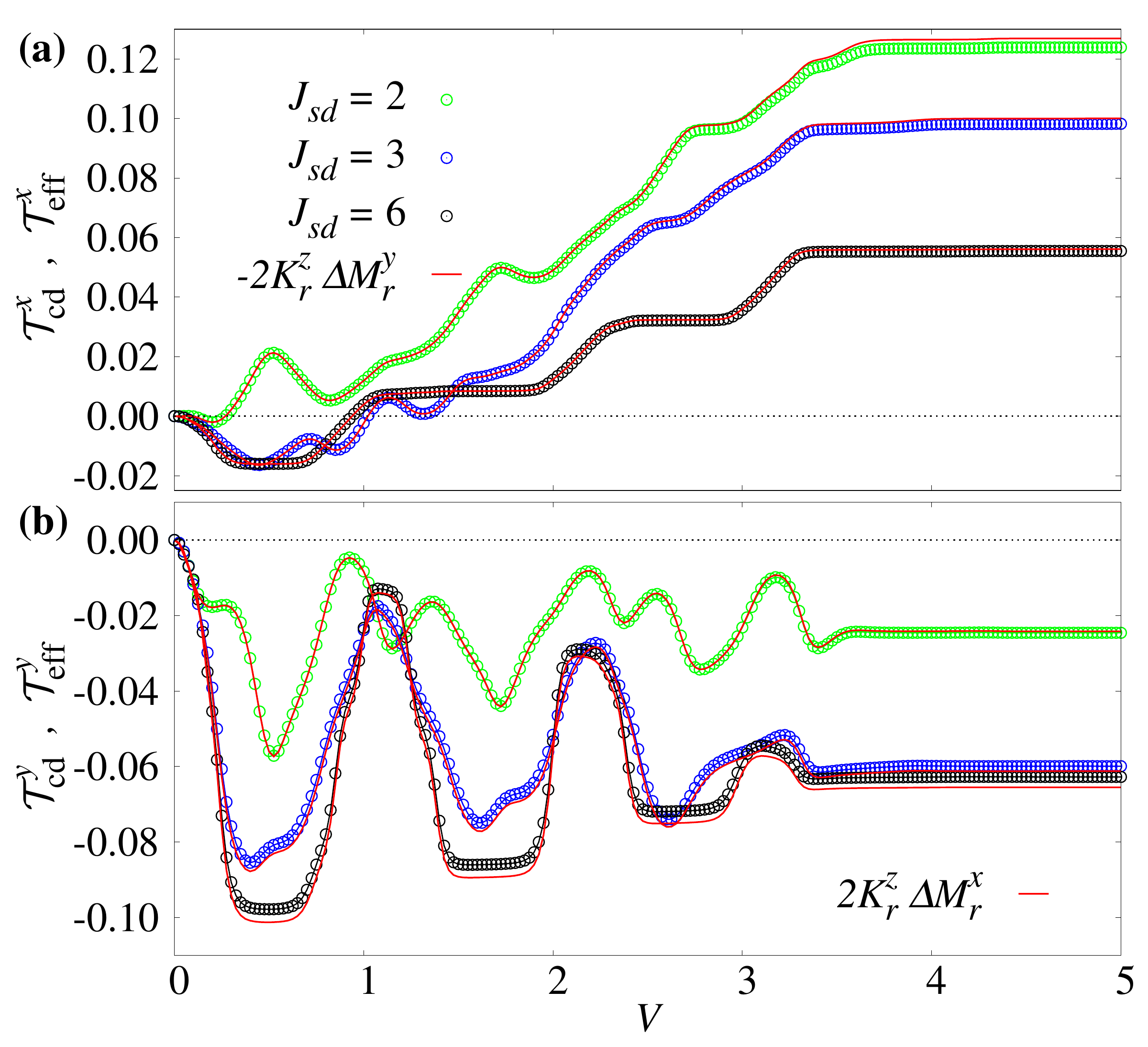}
	\caption{Comparison of the steady state current-driven torques $T_\mathrm{cd}$ and effective local torques 
	$\Teff=2 K_\ell\bm{e}_\ell \times \Delta \bm{M}$ 
	calculated as  function of $V'$ for asymmetric voltage drop.\label{fig:ststL}}
\end{figure}

Because in the symmetric case the only significant tilt of the spins due to the finite voltage is in the $x$-direction and there are no steady state spin currents at the system-lead interfaces, the effective local torques can be further approximated by ${\cal T}_\mathrm{eff}^y = 2 K_r \Delta M_r^x$ (blue circles in Fig.~\ref{fig:ststLR}) for free layer and $2 K_l \Delta M_l^z$ (black crosses in Fig.~\ref{fig:ststLR}) for pinned layer and compared directly to the $\bm{{\cal T}}_\textrm{cd}$ (red lines in Fig.~\ref{fig:ststLR}). 
There is almost a perfect agreement between these three quantities plotted in Fig.~\ref{fig:ststLR} for various $\Jsd$. 


Considering the asymmetric case, the magnetization $\Delta M_r^x$ shown in Fig.~\ref{fig:ststL}(b) differs from the symmetric one in Fig.~\ref{fig:ststLR} mostly in its magnitude. However, the asymmetric case also shows a significant steady state declination for the $y$ component of the magnetization $\Delta M_r^y$. This difference can be again explained by the current-driven torques.
There are finite spin currents flowing between the system and leads. Using the same assumptions as for the symmetric case, we can estimate the effective local torques in the FL to be 
${\cal T}^x_\mathrm{eff} = 2 K_r \Delta M_r^y$ and ${\cal T}^y_\mathrm{eff} = 2 K_r \Delta M_r^x$.
The comparison with the current-driven torque for various $J_{sd}$ (green line in Fig.~\ref{fig:ststLR}) shows again a very good agreement. We can therefore conclude that the finite spin-polarized currents tilt the spins of the free layer not only to the direction of the magnetization in the pinned layer (respective opposite to it), but also perpendicular to both anisotropy fields.   

This, however, opens an interesting question considering the steady state of the symmetric case. There are no spin-polarized steady-state currents flowing between the system and the leads. Yet, the orientation of the classical spins can not be affected by the non-polarized currents. Therefore, the steady state magnetization seems to be fully dictated by the intravalve spin-polarized current $\bm{{\cal J}}_F$. Following the analysis of the transient dynamics, one can conclude that $\bm{{\cal J}}_F$ reflects the polarization of the density of states probed by the chemical potential of the leads. Considering the symmetry of the spectrum as well as the system symmetry and the fact that the charge current plays no role in current-driven torque, there arises a question, if a similar effect can be achieved also in equilibrium.
\begin{figure}[h!]
	\centering
	\includegraphics[width=1.0\columnwidth]{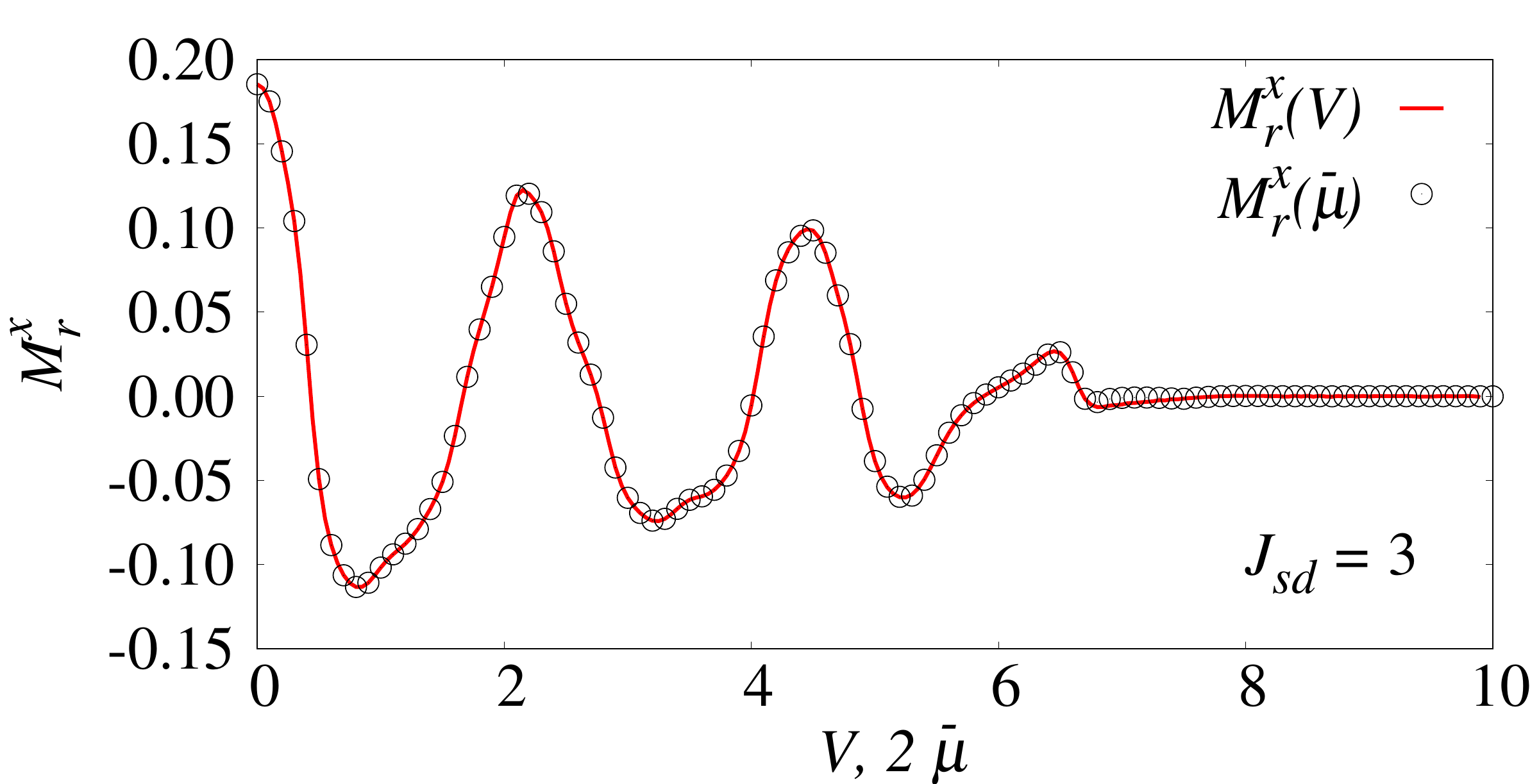}
	\caption{Dependence of steady state orientation of the magnetization $M_r^x$ on symmetric voltage drop $V$ (red line) and the dependence of equilibrium magnetization $M_r^x$ on the spin-valve electrochemical potential $\overline{\mu}$ (circles) for $\Jsd=3$\label{fig:magmu}. The x-axis for the later case is scaled by factor of $2$ to account for $V=2\mu_l=2|\mu_r|$.}
\end{figure} 

This is indeed the case as shown in Fig.~\ref{fig:magmu}. Here we compare the steady state FL magnetization plotted as a function of voltage (line) with its equilibrium counterpart calculated for a fixed $\mu_l=\mu_r=0$ as a function of the electrochemical potential $\overline{\mu}$ for $\Jsd=3$ (circle). They are in perfect agreement. 
For the symmetric voltage drop the effective exchange coupling between the magnetic leads is  defined by the polarization of the steady state DOSh. However, it is not dictated by the Fermi-level of the isolated valve, but by the states probed by the chemical potential of the leads.

\section{Summary}

Magnetic multilayer devices are, besides being important components in a multitude of industrial applications, an ideal tool to investigate various physical concepts. In this paper, we examined the spin-transfer torque and related relaxation processes in a spin-valve system under external voltage bias. To this goal we have adapted a QC-EOM method, which bridges the classical and quantum mechanical approaches by treating the localized spins as classical degrees of freedom that interact with conduction electrons treated as quantum particles. We intentionally focused on regimes where the dynamics of the localized spins can be, to a high degree, represented by the net magnetization in the magnetic layer (macrospin). This allowed us to analyze the numerical results using intuitive approximations. We have shown that even in such idealized cases the dynamics is rather complicated because it reflects a complex relation between the localized spins and conduction electrons.   

In the case of an isolated spin valve, the interplay of classical and quantum degrees of freedom induces a complicated effective exchange interaction $\Jeff$ between the ferromagnetic layers. Although $\Jeff$ is affected mostly by electron states near the Fermi-level, it shows a complex nonmonotonous dependence  on the spin-electron coupling $\Jsd$. 

Coupling the spin valve to metallic leads and introducing a finite bias voltage by shifting their chemical potentials triggers nonequilibrium spin currents, and therefrom spin-transfer torques in the system. Besides influencing the magnetizations in the spin valve, the spin currents also control the relaxation processes of the spin dynamics.  We have observed a resonant character of the relaxation, which is boosted whenever the chemical potential of at least one of the leads matches the maxima in the electronic density of the states of the spin-valve electrons.
 
However, there is a qualitative difference in the transient dynamics, spin relaxation, and even steady state characteristics between the system under symmetric or asymmetric voltage drop with respect to the electrochemical potential of the valve. For example, the relaxation time at high voltages can be of several orders of magnitude longer for the symmetric case than for the asymmetric one. This is a consequence of the fact that there is no long-time spin-polarized current flowing between the system and the leads in the symmetric case. Interestingly, the  steady state magnetization governed by the symmetric voltage drop $V$ can be mapped to a magnetization of the spin valve in equilibrium ($V=0$) with the electrochemical potential adjusted to $\overline{\mu}=V/2$ (where $V$ reflects the voltage drop in the nonequilibrium case).


\section{Acknowledgements}
The authors acknowledge support by the state of Baden-Württemberg through bwHPC
and the German Research Foundation (DFG) through grant no INST 40/467-1 FUGG (JUSTUS cluster). M.\v{Z}. acknowledges support by the Czech Science Foundation via Project No. 22-22419S. This work was supported by the Ministry of Education, Youth and Sports of the Czech Republic through the e-INFRA CZ (ID:90140). The authors thank Richard Koryt\'{a}r and Tom\'a\v{s} Novotn\'y for helpful discussions.

\revappendix
\section{Pad\'e representation}\label[appendix]{sec:AppA}
Besides the integration step the only convergence parameter of the here used QC-EOM is the number of Pad\'e poles in the representation of the Fermi function. 

\begin{figure}[h!]
	\centering
	\includegraphics[width=1.0\columnwidth]{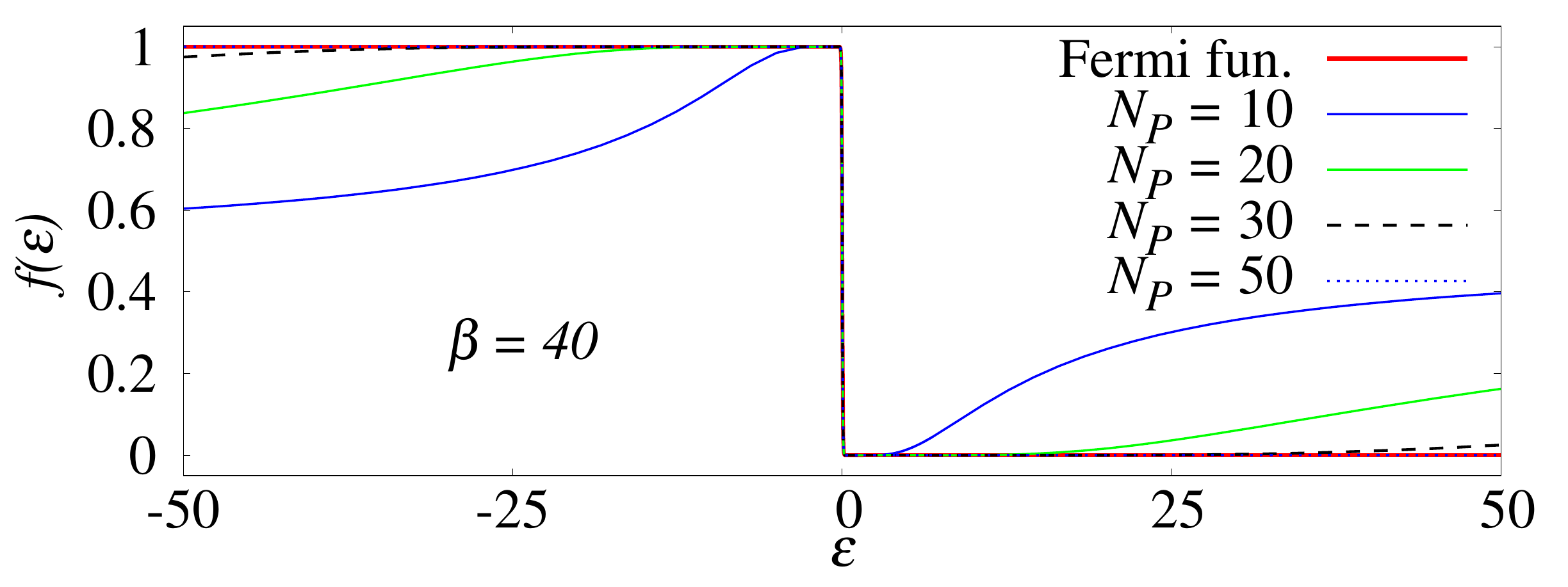}
	\caption{Comparison of exact Fermi function with its approximation Eq.~\eqref{eq:PadeDec} calculated for $N_P=10,20,30,50$. \label{fig:pade}}
\end{figure}

A rather low temperature of $T=0.025$ ($\beta=40$) used in our study requires a relatively high number of Pad\'e poles. The comparison of exact Fermi function and the approximated one, Eq.~\eqref{eq:PadeDec}, with various number of Pad\'e poles is shown in Fig.~\ref{fig:pade}. We have found, that a sufficient precision was obtained in our calculations for $N_P=30$ which we use throughout the paper.  



\section{Localization in one-dimensional valve}
\begin{figure}[!h]
	\centering
	\includegraphics[width=1.0\columnwidth]{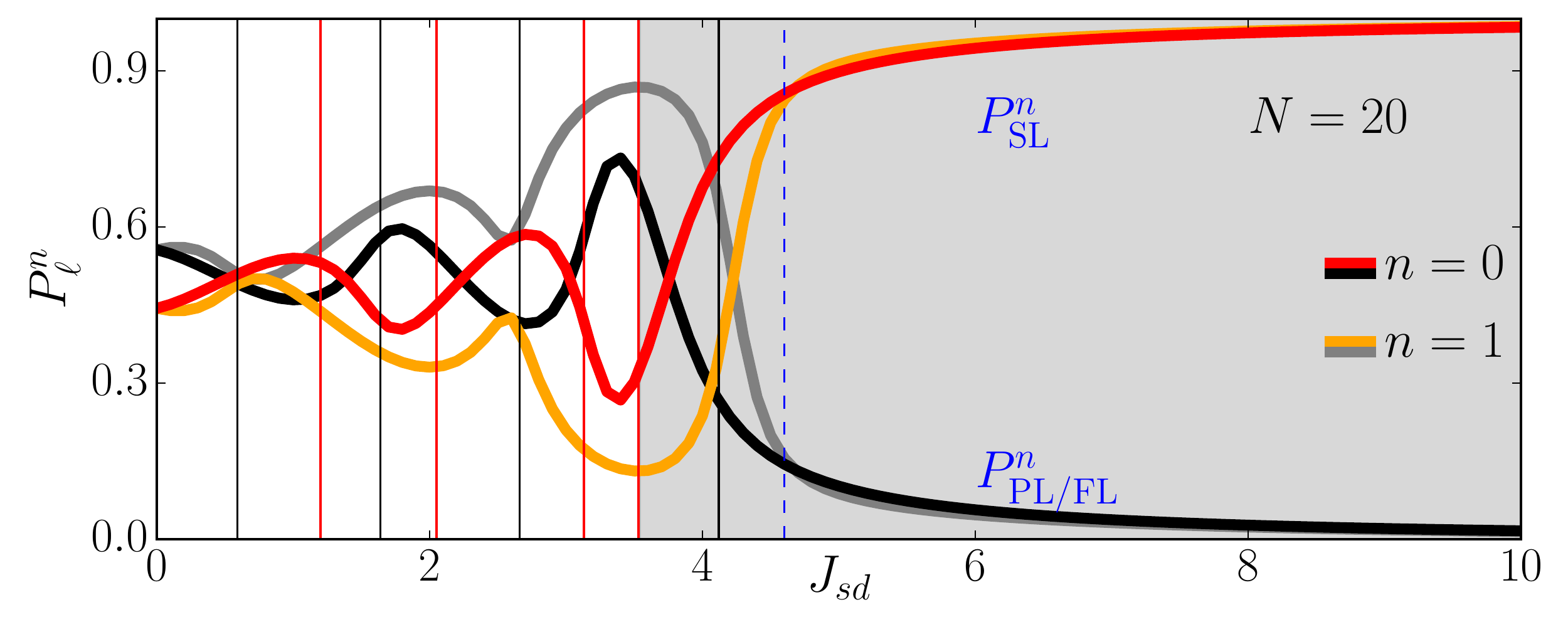}
	\caption{Spatial distribution (local probability) $P_\ell^n$ of energy-eigenstate $\phi_n$ in layer $\ell$, with $P^n_\mathrm{PL/FL}$ summed over all sites $j\in\mathrm{PL/FL}$ (black for $n=0$ and gray for $n=1$), and $P^n_\mathrm{SL}$ over all $j\in\mathrm{SL}$ (red for $n=0$ and orange for $n=1$). 
		Red vertical lines denote $J_{sd}$ for which $\Delta \varepsilon$ has a local maximum, black vertical lines those $J_{sd}$ where $\Delta\varepsilon$ has a local minimum, and gray shaded area denotes the parameter regime in which side-bands split of the main band, as in \cref{fig:AnalysisPeaks}.}\label{fig:App_LocalizationEigenStates}
\end{figure}

\label[appendix]{App:loc}
It is important to note that the regimes of weak ($J_{sd}\lesssim 3$) and strong coupling ($4.5\lesssim J_{sd}$) differ significantly. In the transient regime  $3 \lesssim J_{sd} \lesssim 4.5$ we observe the opening of the gap in the local DOS of the magnetic layers and with it related transition from metallic to insulating like character. In addition, here, the localization of electronic eigenstates starts to change significantly. 
In Fig.~\ref{fig:App_LocalizationEigenStates} we show the spatial distribution $P^n_\ell=\sum_{j\in \ell}\left|\phi_n(\vec{r}_j)\right|^2$, where energy eigenstate $\phi_{n}(\vec{r}_j)$ is the space resolved eigenstate and the sum is restricted to positions within the layer $\ell=\text{FL/PL,SL}$. We focus on the two eigenstates ($n=0,1$) with the lowest eigenenergies, which exhibit the most pronounced localization effects upon increasing $J_{sd}$. For $J_{sd}<3$, it is not possible to infer the typical localization properties of eigenstates from the local probabilities $P^n_\mathrm{L}$. However, there is a significant changes for $J_{sd}>4$, where a clear localization of eigenstates can be observed. The point at which this change occurs, coincides precisely with the point at which spin fluctuations discussed in \cref{sec:SpinValveDynamics} are the strongest.

\section{Rationalization of the macrospin model}\label[appendix]{sec:AppB}

\begin{figure}[!h]
	\centering
	\includegraphics[width=1.0\columnwidth]{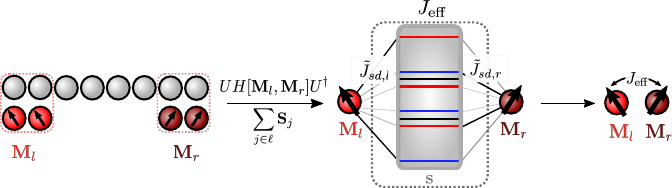}
	\caption{Schematic illustration of the macro-spin approximation: Spin-electron coupled hybrid multi-spin system (left panel) is reduced to two-spin system coupled to the electronic spectrum (middle panel). After integrating out the electronic degrees of freedom, the system is further reduced to a two-spin problem coupled via an effective exchange interaction $J_R$ (right panel).}\label{fig:MappingMulti}
\end{figure}

The macrospin approximation, used in the analysis of the dynamics of the closed spin valve system, can be justified by the scheme illustrated in Fig.~\ref{fig:MappingMulti}. 

In the first step of this approximation we split $\vec{S}_j(t)=\vec{S}_j(0)+\Delta\vec{S}_j(t)$ and diagonalize the single particle Hamiltonian from~\cref{eq:Hq} at $t=0$ resulting in a transformed system 
\begin{align}
	\overline{\bf{H}} &= \sum_\alpha \varepsilon_\alpha \overline{c}_{\alpha}^\dagger \overline{c}_{\alpha}^\pdagger
	+ \Jsd\sum_{\alpha,\alpha'\in \textrm{PL},\textrm{FL}}\overline{c}_{\alpha}^\dagger \overline{S}_{\alpha,\alpha'}(t)^\pdagger \overline{c}_{\alpha'}^\pdagger,\\
	\overline{S}_{\alpha,\alpha'}(t) &= \sum_{j\in \textrm{PL},\textrm{FL}}\sum_{\sigma\sigma'}{\cal U}_{j,\sigma,\alpha}(\bm{\sigma}\cdot\Delta\vec{S}_j(t)^\pdagger)_{\sigma\sigma'}{\cal U}_{j,\sigma',\alpha'},\nonumber
	\label{eq:Hq2}
\end{align}
where ${\cal U}_{j,\sigma',\alpha'}$ are components of the eigenvectors of $\bm{H}(t=0)$. Next, a coarse-grained description of the localized spins is applied, where each magnetic layer is characterized by a local magnetization $\vec{M}_{l,r}=\sum_{j}^{N_{S}}\vec{S}_j$, where $N_{S}$ is the number of spins in a layer ($N_{l(\mathrm{PL})}=N_{r(\mathrm{FL})}=N_S$). We can interpret the simplified system as two macrospins coupled through a spectrum of single particle energies $\varepsilon_\alpha$ via complex time-dependent couplings. Note that using this interpretation one can argue that for low enough temperatures only a few states near the Fermi level will play an important role in the dynamics.   

As a last step in the macro-spin approximation, the central part is approximated by an effective direct exchange coupling $\Jeff$ between the spins, which we assume to be time-independent. Under these assumptions, the problem is reduced to two spins coupled by $\Jeff$.

The exact solution of Eq.~\eqref{eq:5_MeanFieldEOM} used in the extraction of the effective exchange interaction $\Jeff$ between the magnetic layers of the closed spin valve system can be derived by the following steps.   
By recognizing that the cross-product $\vec{M}_{l (r)}\times\vec{M}_{r (l)}$ can be rewritten as a matrix vector multiplication $\vec{A}\times\vec{B}=[\vec{A}]^\times \vec{B}$. Here, $\vec{A},\vec{B}\in\mathds{R}^3$ and $[\vec{A}]^\times=\sum_{\alpha=1}^3 A^\alpha\vec{L}_\alpha$, with $\vec{L}_\alpha$ being the basis of the Lie-algebra $\mathrm{SO}(3)$. These elements generate infinitesimally rotations in $\mathds{R}^3$. and the cross-product in $\mathds{R}^3$ can be expressed using infinitesimal rotations around axis $\vec{A}$
\begin{equation}
	\frac{\dif}{\dif\theta}\bigg|_{\theta=0}\mathcal{R}(\theta,\vec{A})\vec{B}=\vec{A}\times\vec{B}.
\end{equation}
Thus, information about the trajectories of the macrospins can be obtained from infinitesimal rotations.

Each of the macrospins is tracing out a trajectory around the instantaneous position of the other macrospin. Due to the antisymmetry of the cross product, the center of spins is conserved $\vec{M}_l+\vec{M}_r\equiv\vec{M}=\mathrm{const}$. 
Without loss of generality, we assume $\vec{M}=\gamma\hat{\vec{e}}_z$ with $\gamma= M^z_l(t_0)+ M^z_r(t_0)$ determined by the initial condition of both macro-spins because $\partial_t \vec{M}=0$. This assumption is equivalent to a change of the basis into a frame of reference by a rotation of $\theta= -\pi/4$ around the Cartesian $y$-axis in the original frame. 
The trajectory of $\vec{M}_l, \vec{M}_r$ is an intersection between the unit sphere $S^2$ and a straight plane at $z=\gamma$. These constraints are fulfilled by a circle $\mathcal{C}_r \cong S^1$ with radius $r_{\ell}=\sqrt{M_{\ell}^2-\gamma^2}$, where $M_\ell=\vert \vec{M}_\ell\vert^2$. Under these considerations, the general solution is of the form in the rotated frame is

\begin{equation}
	\vec{M}_\ell(t) =\begin{pmatrix}
		r_\ell \cos(\omega t+\phi_\ell)\\
		r_\ell \sin(\omega t+\phi_\ell)\\
		\gamma
	\end{pmatrix}, \qquad \omega = 2\gamma J_{\textrm{eff}}. \label{eq7:MeanField_Solution_app} 
\end{equation}
and thus \cref{eq7:MeanField_Solution} is obtained by rotating \cref{eq7:MeanField_Solution_app} into the original frame of reference by applying the rotation matrix $\mathcal{R}_y(\theta)$ as given in \cref{eq7:MeanField_Solution}.
Due to the symmetry of the system of equations $\omega_l=\omega_r=\omega$, where $\omega$ originates directly from solving \cref{eq:5_MeanFieldEOM}, and the phase difference is exactly $\phi_l-\phi_r=\pi$.

\section{Spin Fluctuations in the intermediate coupling regime} \label{App:SpinFluctuations}
In the main text, we argue that the $J_{sd}$ dependence of the effective coupling $J_\text{eff}$ follows the equilibrium energy difference $\Delta \varepsilon$ between the two highest occupied energy states (\Cref{fig:AnalysisPeaks}). 
However, this correspondence is invalid in the regime $3\lesssim J_{sd}\lesssim4.5$. The purpose of this section is to elucidate the origin of this discrepancy.

\begin{figure}[!h]
	\centering
	\includegraphics[width=1.0\columnwidth]{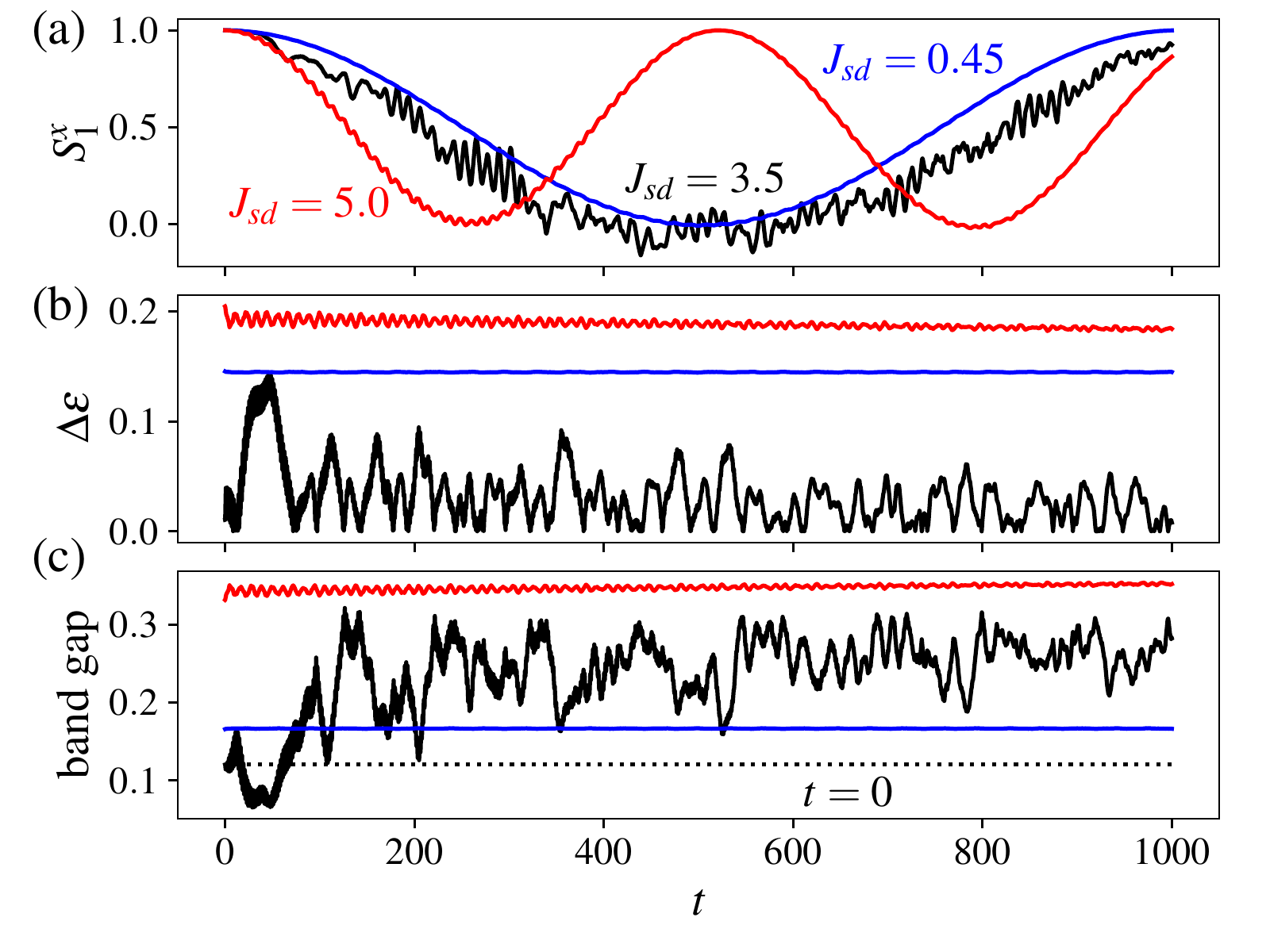}
	\caption{Dynamics of representative classical spin $S_1^x$ at site 1 in the PL (a) and time-dependence of energy-difference $\Delta\varepsilon$ for the same parameter (b) for $J_{sd}=0.45$ (blue) and $J_{sd}=3.5$ (black) in a spin valve with the same system parameters as discussed in \Cref{sec:SpinValveDynamics}.}\label{fig:App_Fluctuations}
\end{figure}

In \Cref{fig:App_Fluctuations} (a) we show the details of the dynamics of a single classical spin in a FM spin valve with the same parameters as in \Cref{sec:SpinValveDynamics} for $J_{sd}=0.45$, $3.5$ and $5$. We diagonalize the Hamiltonian $H(t)\equiv H(\{\vec{S}_i(t)\})$ for each time $t$ to obtain the respective spectrum $\{\varepsilon_n\}$ for the spin configuration $\{\vec{S}_i(t)\}$, and compute therefrom the magnetic splitting $\Delta \varepsilon(t)$ shown in \Cref{fig:App_Fluctuations} (b) and the gap (c). It is obvious that in contrast to the weak and strong coupling regime, both $\Delta \varepsilon(t)$ and the gap show strong fluctuations. In addition, the gap significantly departures from its initial value [dotted black line in (c)]. Note that neither $\Delta \varepsilon(t)$ nor the gap shown in Fig.~\ref{fig:App_Fluctuations} present the actual nonequilibrium values. Nevertheless, they both imply that in contrast to the two other regimes, the static spectrum is insufficient for the analysis of the intermediate regime and so is the macroscopic approximation which assumes $|M_{l,r}|/N_S = 1$.  

\section{Equilibrium spectral properties}\label[appendix]{sec:AppC}

\begin{figure}[!ht]
	\centering
	\includegraphics[width=1.0\columnwidth]{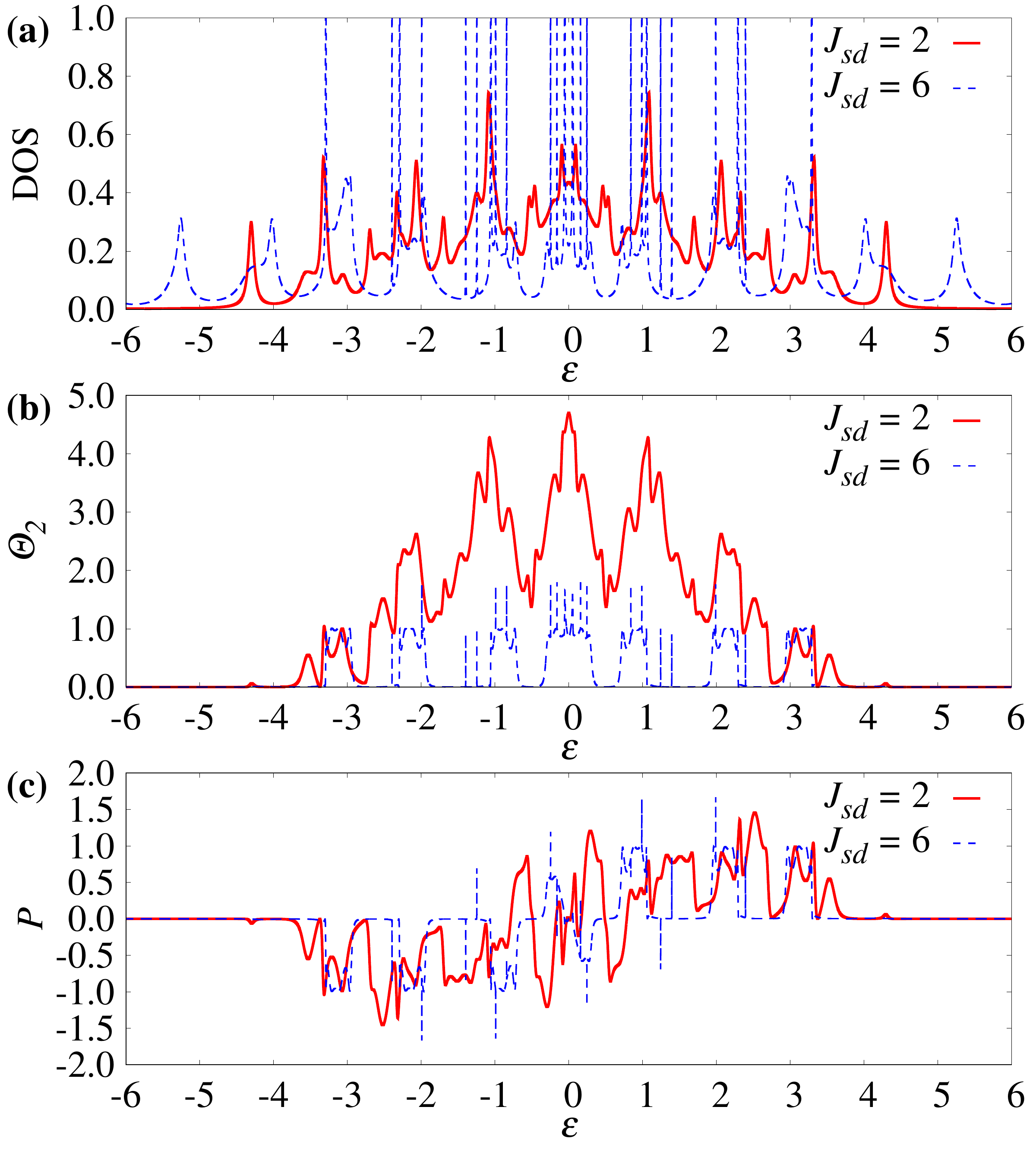}
	\caption{Examples of equilibrium density of states (a), transmission function (b) and spin polarization (c) calculated for the same cases as shown in Fig.~\ref{fig:Mag_td}. \label{fig:eqtrans}}
\end{figure}

The equilibrium spin configuration gives access to the equilibrium DOSh, charge and spin-resolved transmission functions~\eqref{eq:Tran}-\eqref{eq:Dosh} calculated using the equilibrium orientations of the localized spins. In cases where the introduction of the finite voltage leads only to a relatively small reorientation of the classical spins, and therefore a small change of DOSh (e.g., weak interaction $\Jsd$), these equilibrium functions of energy are helpful in the interpretation of some nonequilibrium results. Fig.~\ref{fig:eqtrans} illustrates the equilibrium DOSh and transmission functions calculated for the two cases shown in Fig.~\ref{fig:Mag_td}. 
The sharp states in DOSh for $J_\mathrm{}=6$ in Fig.~\ref{fig:Mag_td}(a) reflect the fast vanishing broadening of the states, originating from the coupling to the leads, in the central part of the system for strong interaction $\Jsd$~\cite{Freericks2004,Freericks2006book,Zonda2019,Zonda2019b}. Related to the strong $\Jsd$ is also the significant drop of transmission when compared with $\Jsd=2$ in Fig.~\ref{fig:Mag_td}(b). This drop reflects the opening of the gap in the magnetic layers and the related change of the character of the spin valve from metallic-like to insulating-like. Note that the states far away from the Fermi level, belonging mostly to the magnetic layers (see Fig.~\ref{fig:SpecTriLayer}), are less relevant for the charge transport from left to right lead than the central ones. Nevertheless, as we discuss later on, they play a role in the relaxation. The spin-polarization of the transmission function measured at the right system interface in Fig.~\ref{fig:Mag_td}(c) shows a rather complicated energy dependence, however, what is important for our analysis is that it is antisymmetric around the Fermi-level.

\section{System with valve-lead interfaces}\label[appendix]{App:AppInt}

%
%

\begin{figure}[!ht]
	\centering     
	\includegraphics[width=1.0\columnwidth]{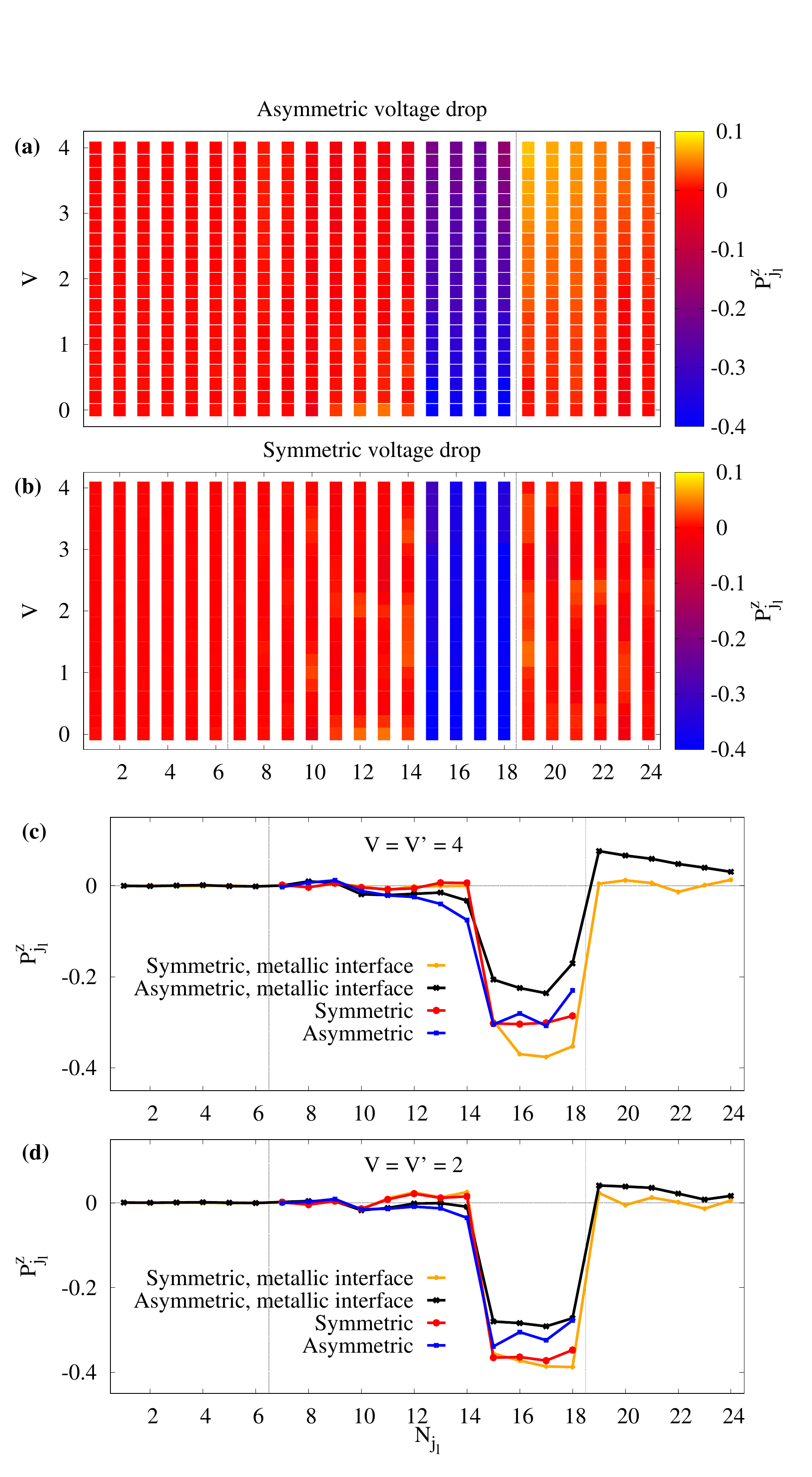}
	\caption{Local electron spin polarization $P^z_{j_l}$ in each monolayer for a valve extended by finite metallic interface with $J_{sd}=2$. The vertical lines show the edges of the bare valve. All other parameters are identical to the system studied in~\cref{sec:Driven}. Panels (a) and (b) show the evolution of $P^z_{j_l}$ with Asymmetric (a) and symmetric (b) voltage drop. Panles (c) and (d) compare $P^z_{j_l}$ of bare and extended valve at $V=V'=4$ an $V=V'=2$. \label{fig:min}}
\end{figure}

In the main text we focus on a simple model where the valve was coupled directly to the semi-infinite leads whose influence on the system is modeled by the current matrices. However, in real systems the surface of the leads can get spin polarized due to the proximity of the magnetic layer and related spin-currents. Therefore,
there arises a question if the differences between the symmetric and asymmetric voltage drop survive such an effect. To partially address this problem we introduce metallic interfaces between the valve and the leads. Basically, the valve is prolonged by six monolayers before the pinned layer and by six metallic monolayers after the free magnetic layer. That way we investigate a valve with $24\times4$ points where PL starts at $N_l=7$ and FL at $N_l=15$ where $N_l$ counts the monolayers from the left edge of the system. In Fig.~\ref{fig:min} we show the local steady state electron spin polarization in $z$-direction $P^z_{j_l}$, i.e., the normalized electron spin-density calculated by summing and normalizing all steady-state contributions in a vertical monolayer
\begin{equation}
	P^z_{j_l}=\sum_{j_v}\textrm{Tr}\,\sigma_z\bm{\rho}_{\{j_l,j_v\}}/\sum_{j_v}\textrm{Tr}\bm{\rho}_{\{j_l,j_v\}},
	\label{eq:Px}
\end{equation}
where $\{j_l,j_v\}$ are longitudinal and vertical coordinates of lattice point $j$. The top two panels show the evolution of $P^z_{j_l}$ with voltage for asymmetric (a) and symmetric (b) voltage drop. The bottom two panels present a comparison of  $P^z_{j_l}$ for systems with and without the finite metallic interface at $V=4$ and $V=2$. The dashed vertical lines mark the edges of the original valve without the finite metallic interface. The tendency towards the polarization of the metallic interface is most visible for the asymmetric voltage drop case at high $V'$ [note the yellow area in panel (a) and the elevation of the black curve at $N_l>18$ in panel (c)]. The effect is most pronounced at the edge of the FL, where it opposes the strong polarization observed within the FL, and vanishes with increasing distance from the FL edge. This effect is, naturally, not captured by the simple model without the interface.  

\begin{figure}[!ht]
	\centering     
	\includegraphics[width=1.0\columnwidth]{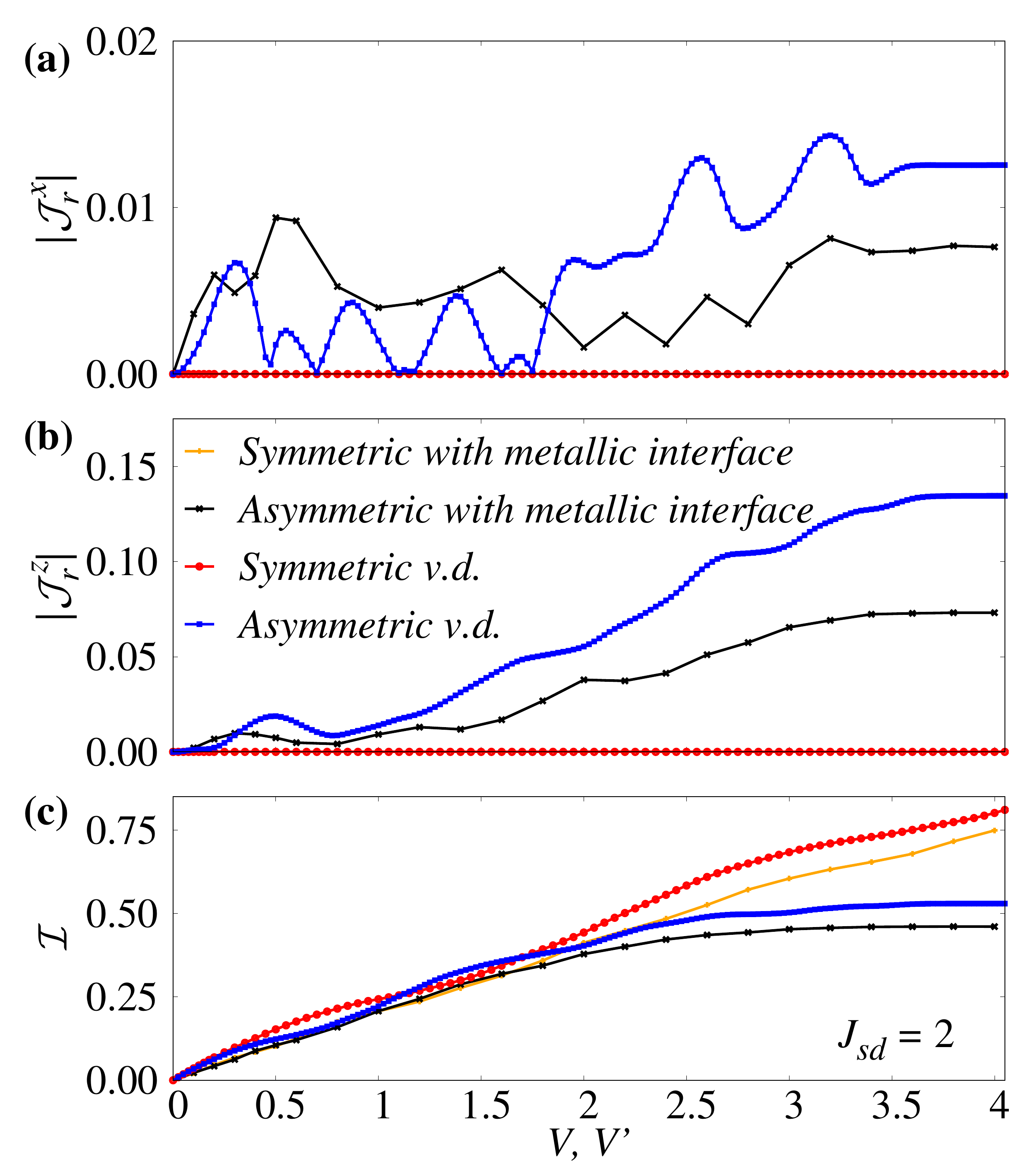}
	\caption{Comparison of the spin (a),(b) and charge (c) currents calculated for a valve without metallic interfaces ($N=12\times 4$, $N_S=4\times 4$) and with finite metallic interface ($N=24\times 4$, $N_S=4\times 4$) for both symmetric and asymmetric voltage drops and $J_{sd}=2$. All other parameters are identical to the system discussed in~\cref{sec:Driven}. \label{fig:iv1}}
\end{figure}

However, when comparing the spin and charge currents measured at the right system-lead interface (\cref{fig:iv1}) we see the same qualitative behavior for the system without finite metallic interface (red and blue lines) and with it (orange and black lines). Note that differences in the course of the current functions are expected. As discussed in the main text, the system is small enough for currents to be sensitive to the energy spectrum of the valve. This is significantly modified by adding the interface which doubles the number of sites of the lattice. Nevertheless, in both cases (with and without the finite interface) there are finite steady-state spin currents for the asymmetric voltage drop and none for the symmetric one. As discussed in the main text this difference is the main reason for the dramatic difference in the spin relaxation of these two cases.

On the other hand, the enlargement of the valve by metallic interfaces seems to broaden the range of voltages at which is the ${\cal I}-V$ characteristic approximately linear. As a consequence, for the extended valve there is a better agreement between the symmetric and asymmetric charge currents for $0.1<V<2$ than for the bare valve. We discuss the linear regime in more detail in the next appendix.      

\section{Linear response regime}\label[appendix]{App:AppLR}
In this appendix we focus on the regime of small voltage. In~\cref{fig:iv} we show the details of {${\cal I}-V$ characteristics calculated for the case of a bare valve with different $J_{sd}$ as well as for a valve extended by the metallic interface discussed in~\cref{App:AppInt}. In all presented cases (and also for various one-dimensional geometries not shown here) the asymmetric and symmetric voltage drops give the same charge current for small voltages if ${\cal I}$ depends approximately linearly on $V$ as expected~\cite{Mahfouzi2013}. For larger voltages, a clear difference appears between symmetric and asymmetric voltage drop. Here we are entering a non-linear regime where the system is more similar to a resonant level model. The transport is sensitive to the complex density of states of the heterostructure.  For example, the current is significantly enhanced whenever the chemical potential of a lead is aligned with a maximum in DOSh.

\begin{figure}[!ht]
	\centering     
	\includegraphics[width=1.0\columnwidth]{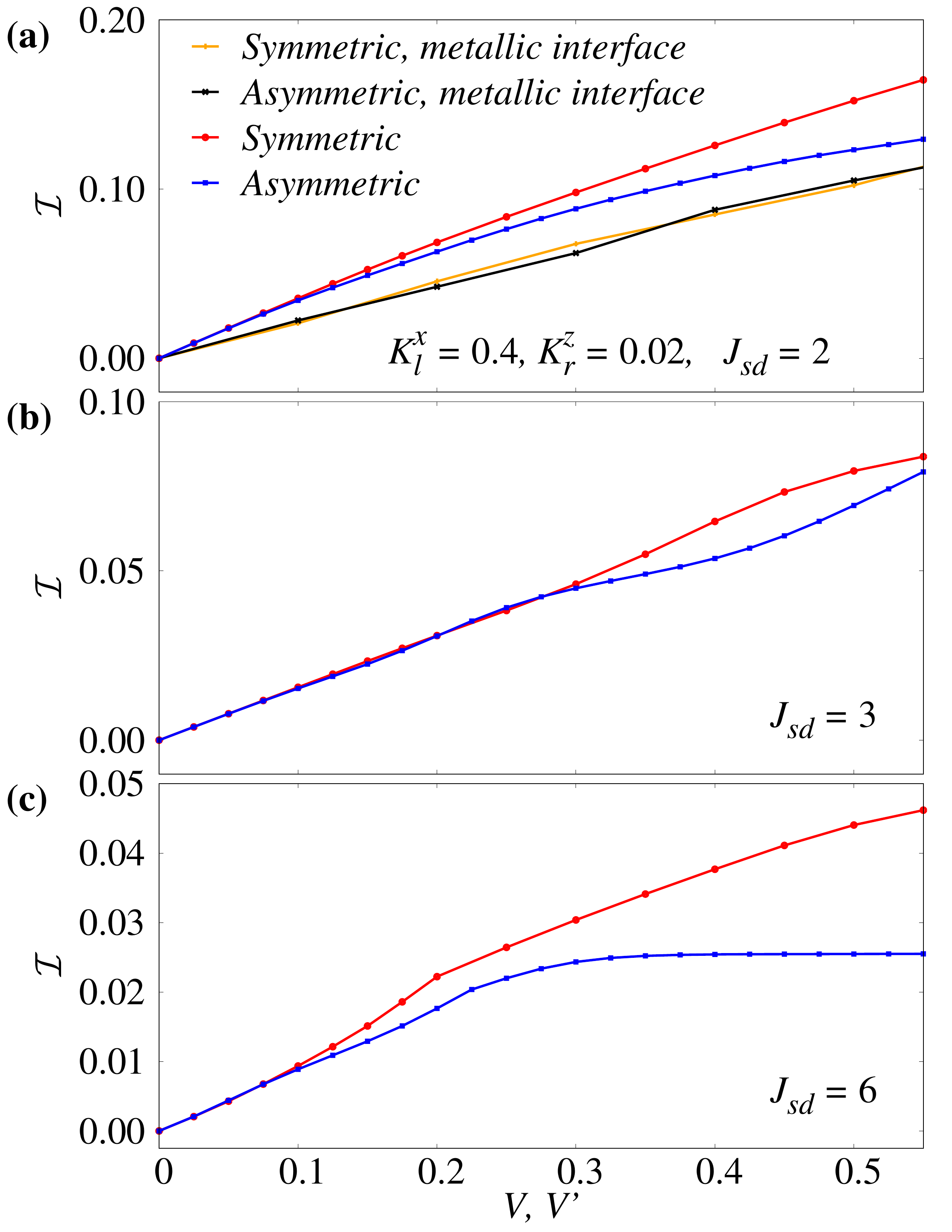}
	\caption{${\cal I}-V$ characteristics at small voltages illustrating that both symmetric and asymmetric voltage drops lead to the same results in the linear response regime observed for $V\ll1$. The orange and black points in panel (a) show the result for a valve extended by an finite metallic interface discussed in~\cref{App:AppInt}. The red and blue data points were calculated for the same parameters as discussed in Sec.~\ref{sec:Driven}. \label{fig:iv}}
\end{figure}

\pagebreak
%

\end{document}